\documentstyle[12pt,epsfig]{article}
\newcommand{\inv}{\left. g^0_A \right|_{\rm inv}}

\newcommand{\Frac}[2]%
{{\textstyle \frac{\mbox{\footnotesize $#1$}\rule[-0.9mm]{0mm}{1mm}}%
{\mbox{\footnotesize $#2$}\rule{0mm}{3.1mm}}}}

\def\cxx{1 - {{4x^{2} P^{2}}\over{Q^{2}}} }
\def\cx{1 - {{2x    P^{2}}\over{Q^{2}}} }

\def\cut{\sqrt{1 - {{4(m^{2}+\lambda^2)}\over {s}} } }

\renewcommand{\thefootnote}{\fnsymbol{footnote}}

\begin{document}
\begin{titlepage}
\vspace*{-12 mm}
\noindent
\begin{flushright}
\begin{tabular}{l@{}}
TUM/T39-99-03 \\
hep-ph/9902280 \\
\end{tabular}
\end{flushright}
\vskip 12 mm
\begin{center}
{\large \bf Constituent quarks and $g_1$
}
\\[14 mm]
{\bf Steven D. Bass}
\footnote[1]{Steven$\_$Bass@physik.tu-muenchen.de} 
\\[10mm]   
{\em Max Planck Institut f\"{u}r Kernphysik, 
Postfach 103 980, \\
D-69029 Heidelberg, Germany}\\[5mm]
{\em Institut f\"ur Theoretische Physik, \\
Physik Department, 
Technische Universit\"at M\"unchen, \\
D-85747 Garching, Germany}\\[5mm]
\end{center}
\vskip 10 mm
\begin{abstract}
\noindent
We review the theory and present status of the proton spin problem 
with
emphasis on the transition between current quarks and constituent 
quarks in QCD.

\end{abstract}
\end{titlepage}
\renewcommand{\labelenumi}{(\alph{enumi})}
\renewcommand{\labelenumii}{(\roman{enumii})}
\renewcommand{\thefootnote}{\arabic{footnote}}
\newpage

\baselineskip=6truemm

\section{Introduction}

Polarised deep inelastic scattering experiments at CERN
\cite{emc,smc,smcfinal,smcqcd},
DESY \cite{hermes} and SLAC \cite{baum,e142,e143a,e154}
have revealed an apparent
two (or more) standard deviations violation of OZI 
in the flavour-singlet axial charge $g_A^{(0)}$
which is extracted from the first moment of $g_1$ 
(the nucleon's first spin dependent structure function).
This discovery
has inspired much theoretical and experimental effort to
understand the internal spin structure of the nucleon.

In this article we review the theory and present status of 
the proton spin problem in QCD.
We start with a simple sum-rule for the spin of the proton 
($+ {1 \over 2}$)
in terms of the angular momentum of its quark and gluonic
constituents:
\begin{equation}
{1 \over 2} \ =
\ {1 \over 2} \Sigma + {\rm L_z} + \Delta g .
\end{equation}
Here, 
${1 \over 2}\Sigma$ and $\Delta g$ are the quark and gluonic 
intrinsic spin contributions to the nucleon's spin and 
${\rm L_z}$ is the orbital contribution.
One would like to understand the spin decomposition, Eq.(1), 
both in terms of the fundamental QCD quarks and gluons and 
also in terms of the constituent quark quasi-particles of low-energy QCD.

In deep inelastic processes the internal structure of the nucleon is
described by the QCD parton model \cite{partal}.
The deep inelastic structure functions may be written as the sum over 
the convolution of ``soft'' quark and gluon parton distributions with 
``hard'' photon-parton scattering coefficients.
The (target dependent) parton distributions describe a flux of quark 
and gluon partons carrying some fraction
$x = p_{+ \rm parton} / p_{+ \rm proton}$
of the proton's momentum
into the hard (target independent) photon-parton interaction which is 
described by the hard scattering coefficients.

In low energy processes the nucleon behaves like a colour neutral system 
of three massive constituent quark quasi-particles interacting 
self consistently with a cloud of virtual pions which is induced by
spontaneous chiral symmetry breaking \cite{cloudy,njlb,njlc}.

One of the most challenging problems in particle physics is to understand 
the transition between the fundamental QCD ``current'' quarks and gluons
and the constituent quarks of low-energy QCD.
The fundamental building blocks are the local QCD quark and gluon 
fields together with the non-local structure \cite{Callan} associated with 
gluon topology \cite{rjc}.

The large mass of the constituent quarks is usually understood 
in terms of dynamical chiral symmetry breaking and 
the interaction of the current quarks with a scalar condensate
(in the Nambu-Jona-Lasinio model \cite{njla,njlb,njlc})
or in terms of scalar confinement (in the Bag model \cite{cloudy}).

Through the axial anomaly in QCD \cite{adler,bell,rjc}, some fraction of 
the spin of the nucleon and of the constituent quark is carried 
by its quark and gluon partons and some fraction is carried by gluon 
topology \cite{bass98}.
The topological winding number is a global property of the theory;
it is independent of the local quark and gluon fields.
When we take the Fourier transform to momentum space 
any 
topological contribution to the nucleon's spin has support only at $x=0$.
This means that whereas 
the nucleon's momentum is given by the sum of the momenta of the partons
\begin{equation}
\sum_{\rm partons} x_{i} \ = \ 1
\end{equation}
one has to be careful about writing equations such as
\begin{equation}
{1 \over 2} \ \stackrel{?}{=}
\ \biggl( {1 \over 2} \Sigma + {\rm L_z} + \Delta g \biggr)_{\rm partons} .
\end{equation}
Some fraction of the nucleon's spin may reside at Bjorken $x=0$.
This effect may be viewed 
as a deformation of the QCD $\theta$ vacuum \cite{Callan} inside a nucleon 
due to tunneling 
processes between vacuum states with different topological winding number.

In semi-classical quark models the quark spin content $\Sigma$ is equal 
to the nucleon's flavour singlet axial charge $g_A^{(0)}$.
Relativistic quark-pion coupling models 
predict
$g_A^{(0)} \simeq 0.6$
with about 40\% of the proton's spin being carried by orbital
angular momentum \cite{schreiber,weise}.

The value of $g_A^{(0)}$ which is extracted from polarised deep inelastic 
scattering experiments is
$g_A^{(0)}|_{\rm pDIS} \simeq 0.2$ -- $0.35$ [1-9];
--- 
significantly less than the semi-classical prediction for $g_A^{(0)}$.

In QCD the axial anomaly induces various gluonic contributions to the
flavour singlet axial charge $g_A^{(0)}$.
Explanations of the small value of $g_A^{(0)}|_{\rm pDIS}$ have 
been proposed based on the QCD parton model \cite{efremov,ar,ccm}
and non-perturbative 
chiral $U_A(1)$ dynamics [25-31].
One finds \cite{efremov,ar,ccm,bass98}
\begin{equation}
g_A^{(0)} 
= 
\Biggl(
\sum_q \Delta q - f {\alpha_s \over 2 \pi} \Delta g \Biggr)_{\rm partons}
+ \ {\cal C}
\end{equation}
where $f$ (=3) is the number of light flavours liberated into
the final state.
Here ${1 \over 2} \Delta q$ and $\Delta g$ are the amount of spin carried 
by quark and gluon partons in the polarised proton and ${\cal C}$ measures 
the gluon-topological contribution to $g_A^{(0)}$ \cite{bass98}.

The topological term ${\cal C}$ has support only at $x=0$; 
it is missed by polarised deep inelastic scattering experiments 
which measure the combination 
$g_A^{(0)}|_{\rm pDIS} = (g_A^{(0)} - {\cal C})$.
If some fraction of the spin of the constituent quark is carried by 
gluon topology in QCD,
then the constituent quark model predictions for $g_A^{(0)}$ 
are not necessarily in contradiction with the small value of 
$(g_A^{(0)} - {\cal C})$
extracted from deep inelastic scattering experiments.
In Section 5 we explain a simple dynamical mechanism for producing 
such an effect.

If there is no topological $x=0$ term, then the small value of 
$g_A^{(0)}|_{\rm pDIS}$ 
would be consistent with the semi-classical prediction
for $\Sigma$ if
$\Delta g$ is both large and positive ($\sim +1.5$ at $Q^2 \simeq 1$GeV$^2$).
If discovered in future experiments, 
such a large $\Delta g$ would pose a challenge for constituent quark models 
which do not naturally include such an effect.

In this article we review our present understanding of the spin structure
of the nucleon,
both at the constituent quark level and in QCD.
We emphasise present and future experiments which could help to unravel
the nucleon's internal spin structure 
by separately measuring the $\Delta q_{\rm parton}$, $\Delta g_{\rm parton}$ 
and ${\cal C}$ contributions to $g_A^{(0)}$.
The structure of the paper is as follows.
In Section 2 we give a brief introduction to deep inelastic scattering 
and the polarised deep inelastic measurements of the nucleon's axial 
charges $g_A^{(k)}$.
In Section 3 we review the theoretical interpretation of the axial 
charges $g_A^{(k)}$ and the 
OZI violation observed in deep inelastic measurements of $g_A^{(0)}$.
The connection between chiral symmetry and the spin structure of the
nucleon is emphasised in Section 3.4.
Sections 4-6 are more technical.
In Section 4 we review the axial anomaly in QCD and its role in our
understanding of the physics of $g_A^{(0)}$.
In Section 5 we discuss the relationship between gluon topology and
the spin structure
of the nucleon.
Section 6 gives an overview of the QCD parton model and its application
to polarised deep inelastic scattering.
In Section 7 we discuss the shape of $g_1$ and the $x$ dependence of 
the various contributions to the nucleon's axial charges and
$\int_0^1 dx\, g_1(x,Q^2)$.
Finally, in Section 8, 
we summarise the information about the spin structure of the 
nucleon which can be learnt in present and future experiments.

We refer to Refs.[32-38]
for further reviews on 
the spin structure of the nucleon in polarised deep inelastic scattering 
and to Ref.\cite{bassdhg}
for a review of polarised photoproduction and the transition to polarised
deep inelastic scattering.

\section{The spin structure function $g_1$}

\subsection{Polarised deep inelastic scattering}

Our present knowledge about the spin structure of the nucleon comes from
polarised deep inelastic scattering experiments.
These experiments involve scattering a high-energy charged lepton beam 
from a nucleon target at large momentum transfer squared.
One measures the inclusive cross-section.
The lepton beam
(electrons at DESY and SLAC and muons at CERN) is longitudinally polarised.
The nucleon target may be either longitudinally or transversely polarised.

Consider polarised $e - p$ scattering.

We work in one photon exchange approximation. 
Whilst the electron photon vertex is described by perturbative QED,
the internal QCD structure of the proton means that 
the photon proton interaction is described in terms 
of various structure functions (form-factors).

Let $p_{\mu}$, $m$ and $s_{\mu}$ denote the momentum, mass and spin of 
the target proton and $q_{\mu}$ denote the momentum of the exchanged 
photon.
Define $Q^2 = -q^2$ and $\nu = p.q$.
Deep inelastic scattering involves working in the Bjorken limit: 
$Q^2$ and $\nu$ both $\rightarrow \infty$ 
with the Bjorken variable $x = {Q^2 \over 2 p.q}$ held fixed.

We specialise to the target rest frame and let $E$ denote the energy of 
the incident electron which is scattered through an angle $\theta$ 
to emerge in the final state with energy $E'$.
Let $\uparrow \downarrow$ denote the longitudinal polarisation of
the beam
and $\Uparrow \Downarrow$ denote a longitudinally polarised proton
target.
The unpolarised and polarised differential cross-sections are:
\begin{equation}
\Biggl(
{d^2 \sigma \uparrow \Uparrow \over d\Omega dE^{'} } +
{d^2 \sigma \uparrow \Downarrow \over d\Omega dE^{'} } 
\Biggr)
=
{8 \alpha^2 (E^{'})^2 \over m Q^4 }
\biggl[ 2  \sin^{2} {\theta \over 2} \ F_1 (x,Q^2) + {m^2 \over \nu}
 \cos^{2} {\theta \over 2} \ F_2 (x,Q^2) \biggr]
\end{equation}
and
\begin{equation}
\Biggl(
{d^2 \sigma \uparrow \Uparrow \over d\Omega dE^{'} } -
{d^2 \sigma \uparrow \Downarrow \over d\Omega dE^{'} } 
\Biggr)
=
{4 \alpha^2 E^{'} \over Q^2 E \nu }
\biggl[ (E+E^{'} \cos \theta ) \ g_1 (x, Q^2) - {2 x m}
\ g_2 (x, Q^2) \biggr] .
\end{equation}
Here $F_1$ and $F_2$ denote the nucleon's first and second 
spin independent structure functions;
$g_1$ and $g_2$ denote the first and second spin dependent 
structure functions.
The structure functions contain all of the target dependent
information in the deep inelastic process.

The unpolarised structure functions have been measured in 
experiments at CERN, DESY, FNAL and SLAC.
We refer to \cite{roberts,deroeck}
for recent reviews of these data and the interpretation of 
$F_1$ and $F_2$.

Polarised deep inelastic scattering experiments with longitudinally 
polarised targets provide the cleanest probe of $g_1$.
In a fixed target experiment 
the contribution of the second spin structure function 
$g_2$ to the differential cross section, Eq.(6),
is suppressed relative to the $g_1$ contribution by 
the kinematic factor ${m \over E}$.
In deep inelastic experiments ${m \over E}$ is typically 
less than 0.03 and the $g_2$ contribution to Eq.(6) 
becomes lost among the experimental errors.
The structure function $g_2$ can be measured using a 
transversely polarised target.  
In this case the two spin structure functions $g_1$ 
and $g_2$ contribute to the spin dependent part of 
the total cross section with equal weight:
\begin{equation}
\Biggl(
{d^2 \sigma \uparrow \Rightarrow \over d\Omega dE^{'} }  -
{d^2 \sigma \uparrow \Leftarrow \over d\Omega dE^{'} } 
\Biggr)
=
{4 \alpha^2 E^{' 2} \over Q^2 E \nu }
\sin \theta
\biggl[ \ g_1 (x, Q^2) + {2 E m \over \nu} \ g_2 (x, Q^2) \biggr] .
\end{equation}

The experimental programme in polarised deep inelastic scattering 
has so far mainly focussed on measurements of $g_1$.
The first measurements of $g_2$ have recently 
been reported by the SMC 
and
SLAC E-143 and E-154 collaborations \cite{exptg2}.
In this review we focus on the physics of $g_1$. 
We refer to Jaffe \cite{jaffeg2} for a theoretical review of $g_2$.

\subsection{The first moment of $g_1$}

When $Q^2 \rightarrow \infty$,
the light-cone operator product expansion relates
the first moment of the structure function $g_1$
to the scale-invariant axial charges of the target nucleon
by \cite{bj,ej,kod,larin}
\begin{eqnarray}
\int_0^1 dx \ g_1^p (x,Q^2) &=&
\Biggl( {1 \over 12} g_A^{(3)} + {1 \over 36} g_A^{(8)} \Biggr)
\Bigl\{1 + \sum_{\ell\geq 1} c_{{\rm NS} \ell\,}
\alpha_s^{\ell}(Q)\Bigr\} \nonumber \\
&+& {1 \over 9} g_A^{(0)}|_{\rm inv}
\Bigl\{1 + \sum_{\ell\geq 1} c_{{\rm S} \ell\,}
\alpha_s^{\ell}(Q)\Bigr\}
\ + \ {\cal O}({1 \over Q^2}).
\end{eqnarray}
Here $g_A^{(3)}$, $g_A^{(8)}$ and $g_A^{(0)}|_{\rm inv}$ are the 
isotriplet, SU(3) octet and scale-invariant 
flavour-singlet axial charges respectively.
The flavour non-singlet $c_{{\rm NS} \ell}$
and singlet $c_{{\rm S} \ell}$ coefficients
are calculable in
$\ell$-loop perturbation theory 
and 
have been calculated to $O(\alpha_s^3)$ precision \cite{larin}.

The first moment of $g_1$ is constrained by low energy 
weak interactions.
For proton states $|p,s\rangle$ with momentum $p_\mu$ and spin $s_\mu$
\begin{eqnarray}
2 m s_{\mu} \ g_A^{(3)} &=&
\langle p,s | 
\left(\bar{u}\gamma_\mu\gamma_5u - \bar{d}\gamma_\mu\gamma_5d \right)
| p,s \rangle _c  \nonumber \\
2 m s_{\mu} \ g_A^{(8)} &=&
\langle p,s |
\left(\bar{u}\gamma_\mu\gamma_5u + \bar{d}\gamma_\mu\gamma_5d
                   - 2 \bar{s}\gamma_\mu\gamma_5s\right)
| p,s \rangle _c
\end{eqnarray}
where the subscript $c$ denotes the connected matrix element.
The isotriplet axial charge $g_A^{(3)}$ is measured independently
in neutron beta decays:
$g_A^{(3)} = 1.267 \pm 0.004$ \cite{datagroup}.
Modulo ${\rm SU_F}(3)$ breaking
\cite{su3},
the flavour octet axial charge
$g_A^{(8)}$ is measured independently in hyperon beta decays:
$g_A^{(8)} = 0.58 \pm 0.03$.
The scale-invariant flavour-singlet axial charge 
$g_A^{(0)}|_{\rm inv}$ 
is defined by \cite{minka}
\begin{equation}
2m s_\mu g_A^{(0)}|_{\rm inv} = 
\langle p,s|
\ E(\alpha_s) J^{GI}_{\mu5} \ |p,s\rangle _c
\end{equation}
where 
\begin{equation}
J^{GI}_{\mu5} = \left(\bar{u}\gamma_\mu\gamma_5u
                  + \bar{d}\gamma_\mu\gamma_5d
                  + \bar{s}\gamma_\mu\gamma_5s\right)_{GI}
\end{equation} 
is the
gauge-invariantly renormalised singlet axial-vector operator 
and
\begin{equation}
E(\alpha_s) = \exp \int^{\alpha_s}_0 \! d{\tilde \alpha_s}\, 
\gamma({\tilde \alpha_s})/\beta({\tilde \alpha_s})
\end{equation}
is a renormalisation group factor 
which corrects 
for the (two loop) non-zero anomalous dimension 
$\gamma(\alpha_s)$ ($= f {\alpha_s^2 \over \pi^2} + {\cal O}(\alpha_s^3)$) 
of $J_{\mu 5}^{GI}$
\cite{koeb,rjc,kod}.
In Eq.(12) $\beta (\alpha_s)$ is the QCD beta function.
We are free to choose the QCD coupling $\alpha_s(\mu)$ at 
either a hard or a soft scale $\mu$.
The singlet axial charge 
$g_A^{(0)}|_{\rm inv}$ 
is independent of the renormalisation scale $\mu$.
It may be measured independently in an elastic 
neutrino proton scattering experiment \cite{kaplan}.

Polarised deep inelastic scattering experiments measure $g_1(x,Q^2)$ 
between some small but finite value $x_{\rm min}$ and an upper value 
$x_{\rm max}$ which is close to one.
Deep inelastic measurements of $g_A^{(3)}$ and $g_A^{(0)}|_{\rm inv}$ 
involve a smooth extrapolation of the $g_1$ data to $x=0$
which is motivated either by Regge theory or by perturbative QCD.
As we decrease $x_{\rm min} \rightarrow 0$
we measure the first moment
\begin{equation}
\Gamma \equiv \lim_{x_{\rm min} \rightarrow 0} \ 
\int^1_{x_{\rm min}} dx \ g_1 (x,Q^2).
\end{equation}
Polarised deep inelastic experiments
cannot, even in principle, measure at $x=0$ with finite $Q^2$.
They miss any possible $\delta (x)$ terms which might exist in 
$g_1$ at large $Q^2$.

Assuming no isotriplet $\delta (x)$ term in $g_1$, 
polarised deep inelastic scattering experiments at CERN 
\cite{emc,smc,smcfinal,smcqcd},
DESY \cite{hermes} and SLAC \cite{e143a,e154} 
have verified 
the Bjorken sum-rule \cite{bj}
\begin{equation}
I_{Bj} = 
\int_0^1 dx \Biggl( g_1^p - g_1^n \Biggr)
=
\frac{g_A^{(3)}}{3} 
\left[1 - \frac{\alpha_s}{\pi} - 3.58 \left(\frac{\alpha_s}{\pi} \right)^2 
        - 20.21 \left(\frac{\alpha_s}{\pi} \right)^3 \right] 
\end{equation}
for the isovector part of $g_1$
to 10\% accuracy.
They have also revealed an apparent two standard deviations 
violation 
of OZI in the flavour singlet axial charge
extracted from polarised deep inelastic scattering:
\begin{equation}
\left. g^{(0)}_A \right|_{\rm pDIS} = 0.2 - 0.35. 
\end{equation}
This number compares with 
$g_A^{(8)} = 0.58 \pm 0.03$ 
from hyperon beta-decays \cite{su3}.

The small $x$ extrapolation of $g_1$ data
is presently the largest source of experimental error on measurements 
of the nucleon's axial charges from deep inelastic scattering.

\section{Interpretation of $g_A^{(k)}$}

The small value of $g_A^{(0)}|_{\rm pDIS}$ measured in polarised 
deep inelastic scattering
has inspired many theoretical ideas about the spin structure of
the nucleon.
The original EMC measurement \cite{emc} of $g_A^{(0)}|_{\rm pDIS}$ 
came as a surprise since, in the pre-QCD parton model,
$g_A^{(0)}$
is interpreted as the fraction of the proton's spin which 
is carried
by the spin of its quarks 
--- and since the original EMC measurement was consistent 
with zero!
It is amusing to speculate how QCD might have developed if 
that measurement
had been available and current at the time 
when
Gell-Mann and collaborators discovered the Eightfold Way symmetry.

How should we interpret the axial charges $g_A^{(k)}$ in QCD ?

Define
\begin{equation}
2m s_{\mu} \Delta q 
= 
\langle p,s | 
\biggl( {\overline q} \gamma_{\mu} \gamma_5 q \biggr)_{GI}
| p,s \rangle _c .
\end{equation}
The axial charges may be written
\begin{eqnarray}
g_A^{(3)} &=& \Delta u - \Delta d \\ \nonumber
g_A^{(8)} &=& \Delta u + \Delta d - 2 \Delta s \\ \nonumber
g_A^{(0)} \equiv 
g_A^{(0)}|_{\rm inv}/E(\alpha_s) &=& \Delta u + \Delta d + \Delta s .
\end{eqnarray}
The non-singlet axial charges are scale invariant. 
The flavour-singlet combination 
$g_A^{(0)} = \Delta u + \Delta d + \Delta s$ 
depends on the renormalisation scale $\mu$; 
it evolves with the two loop anomalous dimension $\gamma(\alpha_s)$.
The scale dependent $g_A^{(0)}(\mu^2)$ is frequently used in
theoretical descriptions of deep inelastic scattering 
where it is common to set the renormalisation scale $\mu^2$ 
equal to the virtuality $Q^2$ of the hard photon, 
make a perturbative expansion of $E(\alpha_s)$ and then absorb $E(\alpha_s)$
into
the singlet Wilson coefficient 
$\Bigl\{1 + \sum_{\ell\geq 1} c_{{\rm S} \ell\,} \alpha_s^{\ell}(Q)\Bigr\}$.
While this is a legitimate theoretical procedure for describing
the first moment of $g_1$ at large $Q^2$
it is important to bear in mind that
physical observables do not depend on the theorist's choice
of renormalisation scale.
$g_A^{(0)}|_{\rm inv}$ is a physical observable whereas the 
renormalisation-scale dependent $g_A^{(0)}(\mu^2)$ is not
\cite{rjcthanks}.

In the rest of this Section we discuss the interpretation of the axial 
charges $g_A^{(k)}$ in constituent quark models (Section 3.1) and in 
QCD (Sections 3.2 and 3.3).
In Section 3.4 we highlight the relation between chiral symmetry and
the spin structure of the nucleon.

\subsection{Constituent quarks and $g_A^{(k)}$}

In semi-classical quark models $\Delta q$ is interpreted as the amount
of spin carried by quarks and antiquarks of flavour $q$,   there is no 
$E(\alpha_s)$ factor and $\Sigma = g_A^{(0)}$.
Relativistic quark-pion coupling models such as the Cloudy Bag 
\cite{cloudy}
which 
contain no explicit strange quark or gluon degrees of freedom
\footnote
{The gluonic degrees of freedom are integrated out into the confinement 
 potential.}
predict $g_A^{(0)} = g_A^{(8)}$.
Whilst these models do not explain the small value 
of $g_A^{(0)}|_{\rm pDIS}$
they do give a good account of the 
flavour non-singlet axial charges $g_A^{(3)}$ and $g_A^{(8)}$.

First, consider the static quark model.
The simple SU(6) 
proton wavefunction 
\begin{eqnarray}
|p \uparrow \rangle = 
{1 \over \sqrt{2}} | u \uparrow (ud)_{S=0} \rangle
&+&
{1 \over \sqrt{18}} | u \uparrow (ud)_{S=1} \rangle
-
{1 \over 3} | u \downarrow (ud)_{S=1} \rangle \\ \nonumber
-
{1 \over 3} | d \uparrow (uu)_{S=1} \rangle
&+&
{\sqrt{2} \over 3} | d \downarrow (uu)_{S=1} \rangle
\end{eqnarray}
yields
$g_A^{(3)} = {5 \over 3}$ and $g_A^{(8)}=g_A^{(0)}=1$.

In relativistic quark models one has to take into account
the four-component Dirac spinor
$\psi = \biggl({ f \atop \sigma .{\hat{r}} g }\biggr)$.
The lower component of the Dirac spinor is p-wave with intrinsic spin 
primarily pointing in the opposite direction to spin of the nucleon.
Relativistic effects renormalise the NRQM axial charges 
by a factor $(f^2 - {1 \over 3} g^2)$
with a net transfer of angular momentum 
from intrinsic spin to orbital angular momentum.
In the MIT Bag $(f^2 - {1 \over 3} g^2) = 0.65$
reducing 
$g_A^{(3)}$ from ${5 \over 3}$ to 1.09.
Centre of mass motion then increases the axial 
charges by 
about 20\% 
bringing $g_A^{(3)}$ close to its physical value 1.27
\footnote{
The renormalisation of $g_A^{(3)}$ from ${5 \over 3}$ 
to $\simeq 1.25$
is also found in light-cone binding models \cite{schlumpf}.}
.

The pion cloud of the nucleon also renormalises the nucleon's axial 
charges by shifting intrinsic spin into orbital angular momentum.
Consider the Cloudy Bag Model (CBM).
Here,
the Fock expansion of the nucleon in terms of a bare MIT nucleon 
$|{\rm N}\rangle$ and baryon-pion $|{\rm N} \pi \rangle$ 
$|{\Delta \pi}\rangle$ Fock states converges rapidly.
We may safely truncate the Fock expansion at the one pion level.
The CBM axial charges are \cite{schreiber}:
\begin{eqnarray}
g_A^{(3)} &=& 
{5 \over 3} {\cal N}
\biggl( 1 - {8 \over 9}  P_{N \pi} - {4 \over 9} P_{\Delta \pi}
            + {8 \over 15} P_{N \Delta \pi} \biggr) \\ \nonumber
g_A^{(0)} &=& 
\ \ \ {\cal N}
\biggl(1 - {4 \over 3}  P_{N \pi} + {2 \over 3} P_{\Delta \pi} \biggr) .
\end{eqnarray}
Here, 
${\cal N}$ takes into account the relativistic factor 
$(f^2 - {1 \over 3} g^2)$ 
and centre of mass motion in the Bag.
The coefficients
$P_{N \pi}=0.2$ and $P_{\Delta \pi}=0.1$ 
denote the probabilities to find the physical nucleon in 
the 
$|N \pi \rangle$ and $|\Delta \pi \rangle$ Fock states respectively
and $P_{N \Delta \pi}=0.3$ is the interference term. 
The bracketed pion cloud renormalisation factors in Eq.(19) 
are 0.94 for $g_A^{(3)}$ and 0.8 for $g_A^{(0)}$.
Through the Goldberger-Treiman, the small pion cloud renormalisation 
of $g_A^{(3)}$ translates into a small pion cloud renormalisation of 
$g_{\pi NN}$,
which is necessary to treat the pion cloud in low order perturbation 
theory \cite{cloudy}.
With a 20\% centre of mass correction,
the CBM predicts 
$g_A^{(3)} \simeq 1.25$ and $g_A^{(0)} = g_A^{(8)} \simeq 0.6$.
Similar numbers \cite{weise} are obtained in the Nambu-Jona-Lasinio model.

Including kaon loops into the model generates a small $\Delta s$
$\simeq -0.003$ \cite{henley} in the Cloudy Bag Model 
and $-0.006$ \cite{weise} in the Nambu-Jona-Lasinio model.
These values are an order of magnitude smaller than the value of
$\Delta s$
extracted from polarised deep inelastic scattering experiments by 
combining Eq.(15) with $g_A^{(8)}$: 
$\Delta s$ between -0.13 and -0.07.

In QCD the axial anomaly induces various gluonic contributions
to $g_A^{(0)}$.
Since gluons are flavour singlet, 
it follows that, modulo flavour SU(3) breaking, 
explicit 
gluonic contributions to $\Delta q$ will cancel 
in the isotriplet and SU(3) octet axial charges $g_A^{(3)}$ and $g_A^{(8)}$.

\subsection{The renormalisation group factor $E(\alpha_s)$}

The first QCD effect that we consider is the renormalisation group factor 
$E(\alpha_s)$.
The scale dependent axial charge $g_A^{(0)}$ is related to the scale 
invariant 
$g_A^{(0)}|_{\rm inv}$ 
by 
$g_A^{(0)} = g_A^{(0)}|_{\rm inv}/E(\alpha_s)$.

One popular idea \cite{grv,bagsf} is that the physics of confinement 
and dynamical symmetry breaking determines the parton 
distributions at some 
low scale $\mu_0^2 \sim 0.3$GeV$^2$.
Parton distributions may be calculated at the scale $\mu_0^2$ 
using one's favourite quark model, evolved using perturbative 
QCD to deep inelastic values of  $Q^2$ and then compared with data.
In this approach it is natural to associate the quark model
predictions of $g_A^{(0)}$ 
with 
$g_A^{(0)}(\mu_0^2)$ 
instead of the scale-invariant quantity $g_A^{(0)}|_{\rm inv}$ in QCD.

How big is $E(\alpha_s)$ ?

The perturbative QCD expansion of $E(\alpha_s)$ is
\begin{eqnarray}
E(\alpha_s) &=& 
  1  - {24 f \over 33 - 2f} {\alpha_s \over 4\pi} 
  \\ \nonumber
  &+& 
  {1 \over 2}
  \biggl( { \alpha_s \over 4 \pi } \biggr)^2
  {f \over 33 - 2f }
  \biggl( {16f \over 3} - 472 + 72 {102 - {14f \over 3} \over 33 - 2f}
  \biggr)
  + {\cal O}(\alpha_s^3)
\end{eqnarray}
where $f$ is the number of flavours.
To ${\cal O}(\alpha_s^2)$ the perturbative expansion (20) remains close 
to one -- even for large values of $\alpha_s$.
If we take $\alpha_s \sim 0.6$ as typical of the infra-red 
\footnote{
The coupling $\alpha_s (\mu_0^2) \simeq 0.6$ is the optimal model input
which is found in both the GRV \cite{grv} and Bag Model \cite{bagsfa} 
fits to deep inelastic structure function data.
It is interesting to note that this is the same coupling where Gribov 
\cite{gribov}
suggested that perturbative QCD should give way to something approaching 
a constituent-quark pion coupling model.}
then
\begin{equation}
E(\alpha_s) \simeq
1 - 0.13 - 0.03 = 0.84.
\end{equation}
Here -0.13 and -0.03 
are the ${\cal O}(\alpha_s)$ and ${\cal O}(\alpha_s^2)$ 
corrections respectively.
Perturbative QCD evolution is insufficient to reduce the flavour-singlet 
axial-charge from its naive value 0.6 to the value (15) extracted from 
polarised deep inelastic scattering.

\subsection{Gluons and $g_A^{(0)}$}

In QCD the axial anomaly \cite{adler,bell} induces various gluonic 
contributions to $g_A^{(0)}$.  
One finds \cite{efremov,ar,ccm,bass98}
\begin{equation}
g_A^{(0)} 
= 
\Biggl(
\sum_q \Delta q - f {\alpha_s \over 2 \pi} \Delta g \Biggr)_{\rm partons}
+ \ {\cal C}
\end{equation}
where $f$ (=3) is the number of light-quark flavours.
Here
$\Delta q_{\rm parton}$ and $\Delta g_{\rm parton}$ 
are interpreted as the amount of spin carried by 
quark and gluon 
partons in the polarised nucleon
and ${\cal C}$ measures 
any contribution to $g_A^{(0)}$ from gluon topology 
\cite{bass98}.
In leading order QCD evolution $\Delta g_{\rm parton}$ 
evolves as $1/\alpha_s$ 
so the product 
$-{\alpha_s \over 2 \pi} \Delta g_{\rm parton}$ scales 
at very large $Q^2$ \cite{ar}.

The three terms in Eq.(22) are separately measurable. 
We now outline the physics associated with each of 
these three contributions 
--- 
a detailed discussion is given in Sections 4-6 below.

The QCD parton model \cite{partal} describes $g_1$ at finite $x$ 
(greater than zero).
The polarised gluon contribution to Eq.(22) is characterised by 
the contribution to the first moment of $g_1$ from 
two-quark-jet events 
carrying large transverse momentum squared
$k_T^2 \sim Q^2$ \cite{ccm}
which are generated by photon-gluon fusion --- see Section 6.
The polarised quark contribution $\Delta q_{\rm parton}$ 
is associated with the first moment of the measured $g_1$ 
after these two-quark-jet events are subtracted from the total data set.

The term ${\cal C}$ measures any topological contribution to $g_A^{(0)}$ 
and has support only at $x=0$.
Suppose that gluon topology
contributes an amount
${\cal C}$ to the flavour-singlet axial charge $g_A^{(0)}$.
The flavour-singlet axial charge
which is 
extracted from a polarised deep inelastic experiment is
$(g_A^{(0)} - {\cal C})$.
In contrast, elastic ${\rm Z}^0$ exchange processes such as
$\nu p$ elastic scattering \cite{garvey} 
and parity violation in light atoms \cite{parity,nachtmann} 
measure the full $g_A^{(0)}$ \cite{kaplan}.
One can, in principle, measure the topology term ${\cal C}$
by comparing the flavour-singlet axial charges
which are extracted from polarised deep inelastic and
$\nu p$ elastic scattering experiments.

{\it If}
some fraction of the spin of the constituent quark is carried by 
gluon topology in QCD,
then the constituent quark model predictions for $g_A^{(0)}$ are
not necessarily in contradiction with the small value of 
$(g_A^{(0)} - {\cal C})$
which is extracted from deep inelastic scattering experiments.
The Ellis-Jaffe conjecture \cite{ej} 
($g_A^{(8)} \simeq g_A^{(0)}|_{\rm inv}$)
may hold in constituent quark models and in QCD but fail if we 
consider only the partonic ($x>0$) contributions to the nucleon's
axial-charges.
\footnote{
Possible $\delta(x)$ terms in deep inelastic structure functions 
have also been discussed within the context of Regge theory
where they are induced by Regge fixed poles with non-polynomial
residue \cite{broadhurst}.}

In the absence of the topological term (${\cal C} =0$),
the small value of $g_A^{(0)}$ extracted from polarised 
deep inelastic scattering
would be consistent with the semi-classical predictions 
for $\Sigma$ 
{\it if}
$\Delta g_{\rm parton}$ is both large and positive
($\sim +1.5$ at $Q^2 \simeq 1$GeV$^2$).
At the same time, such a large $\Delta g_{\rm parton}$ 
would pose a challenge for 
constituent quark models, which do not naturally include such an effect.
The size of $\Delta g_{\rm parton}$ is one of the key issues in QCD spin 
physics at the present time --- see Section 6.3.

\subsection{Chiral symmetry and $g_A^{(k)}$}

We have seen in Sections 3.1-3.3 that the axial charges $g_A^{(k)}$ 
measure the partonic spin structure of the nucleon.
The isotriplet Goldberger-Treiman relation \cite{dashen}
\begin{equation}
2 m g_A^{(3)} = f_{\pi} g_{\pi NN}
\end{equation}
relates $g_A^{(3)}$ and therefore $(\Delta u - \Delta d)$ 
to the product of the pion decay constant $f_{\pi}$ and the 
pion-nucleon coupling constant $g_{\pi NN}$.
This result is non-trivial.
It means that the spin structure of the nucleon measured in 
high-energy, high $Q^2$ polarised deep inelastic scattering
is intimately
related
to spontaneous chiral symmetry breaking and low-energy pion physics.

Isosinglet extensions of the Goldberger-Treiman relation 
are quite subtle because of the $U_A(1)$ problem whereby 
gluonic
degrees of freedom mix with the flavour-singlet 
Goldstone 
state
to increase the masses of the $\eta$ and $\eta'$ mesons \cite{fmink}.
If we work in the approximation $m_u = m_d$, 
then the $\eta - \eta'$ mass matrix becomes \cite{christos}
\begin{equation}
M^2_{\eta - \eta'} =\
\left(\begin{array}{cc} 
{4 \over 3} m_{\rm K}^2 - {1 \over 3} m_{\pi}^2  &
- {2 \over 3} \sqrt{2} (m_{\rm K}^2 - m_{\pi}^2) \\
\\
- {2 \over 3} \sqrt{2} (m_{\rm K}^2 - m_{\pi}^2) &
{2 \over 3} m_{\rm K}^2 + {1 \over 3} m_{\pi}^2 + \chi(0)/N_c
\vphantom{\inv}  
\end{array}\right) .
\end{equation}
Here we work in the $(\lambda_8, \lambda_0)$ 
basis.
The gluonic contribution to the mass of the flavour singlet 
state is
$\chi(0)/N_c$
where
$\chi(0)$ is the topological susceptibility 
\cite{christos,witten}
and $N_c$ is the number of colours in QCD. 
We diagonalise the matrix (24) to obtain 
the masses
of the physical $\eta$ and $\eta'$ mesons:
\begin{equation}
m^2_{\eta', \eta} = (m_{\rm K}^2 +\chi(0)/2N_c) 
\pm {1 \over 2} 
\sqrt{(2 m_{\rm K}^2 - 2 m_{\pi}^2 - \chi(0)/3 N_c)^2 
   + {8 \over 9} \chi(0)/N_c^2} .
\end{equation}
If we turn off the gluon mixing term, 
then one finds
$m_{\eta'} = \sqrt{2 m_{\rm K}^2 - m_{\pi}^2}$ 
and
$m_{\eta} = m_{\pi}$.
The best fit to the $\eta$ and $\eta'$ masses 
from the quadratic mass formula 
(25) is
$m_{\eta} = 499$MeV and $m_{\eta'} = 984$MeV
corresponding to 
taking $\chi(0)/N_c = 0.73$GeV$^2$ 
and an $\eta - \eta'$ mixing angle $\theta \simeq 18.2$ degrees.
The physical masses are $m_{\eta} = 547$MeV and $m_{\eta'} = 958$MeV.
Several explanations of the $U_A(1)$ problem and the dynamical
origin of the 
$\chi(0)/N_c$ term have been proposed based on instantons 
\cite{thooft,thrept}
and
large $N_c$ arguments \cite{witten,mink}.
The axial anomaly is central to each of these explanations.

Working in the chiral limit, Shore and Veneziano \cite{venez} 
have used the low-energy $U_A(1)$ effective action of QCD to 
derive the flavour-singlet Goldberger-Treiman relation
\begin{equation}
2m g_A^{(0)} = \sqrt{ \chi' (0)} g_{\phi_0 NN} .
\end{equation}
Here $\phi_0$ is the flavour-singlet Goldstone boson which would exist 
in a gedanken world where OZI is exact --- 
for example, 
taking $N_c$ to infinity in Eqs.(24,25) 
with $\chi(0)$ held constant as a function of $N_c$ \cite{witten}.
The $\phi_0$ 
is a theoretical object and not a physical state in the spectrum.
$\chi'(0)$ is the first derivative of the topological susceptibility.
viz.
\begin{equation}
\chi'(0) = 
\lim_{k^2 \rightarrow 0} {{\rm d} \over {\rm d}k^2}
\biggl( 
\int {\rm d}z
e^{ik.z} i \langle {\rm vac} | T Q(z) Q(0) | {\rm vac} \rangle
\biggr) 
\end{equation}
where
$Q(z) = {\alpha_s \over 2 \pi} G_{\mu \nu}{\tilde G}^{\mu \nu} (z)$
is the 
topological charge density.
The value of $g_A^{(0)}$ which appears in the flavour-singlet 
Goldberger-Treiman relation (26) 
includes any contribution 
from gluon topology at Bjorken $x$ equal to zero.

The important feature of Eq.(26) is that $g_A^{(0)}$ factorises 
into the 
product of the target dependent coupling $g_{\phi_0 NN}$ 
and the target
independent susceptibility term $\sqrt{\chi'(0)}$.
The scale dependence of $g_A^{(0)}$ is carried by 
$\sqrt{ \chi' (0)}$; 
the coupling $g_{\phi_0 NN}$ is scale independent.
Motivated by this observation,
Narison, Shore and Veneziano \cite{narison} conjectured that any OZI 
violation in $g_A^{(0)}|_{\rm inv}$ 
might be carried by the target independent 
factor $\sqrt{ \chi' (0)}$ and that $g_{\phi_0 NN}$ might be free of 
significant OZI violation.
In a different approach, 
Brodsky, Ellis and Karliner \cite{bek} have used a particular version 
of the Skyrme model to argue that $g_{\phi_0 NN}$ might be OZI suppressed.
The target (in-)dependence of the OZI violation in $(g_A^{(0)}-{\cal C})$
may be tested in semi-inclusive measurements of polarised deep inelastic 
scattering in the target fragmentation region \cite{gsven}. 
These experiments \cite{ozidep} could be performed with a polarised 
proton beam at HERA \cite{albert}.

The flavour-singlet Goldberger-Treiman relation (26) can also 
be written 
as the sum of two terms involving the coupling of the physical
$\eta'$ and a pseudoscalar ``glueball'' object, $G$, 
to the nucleon \cite{venez}:
\begin{equation}
2m g_A^{(0)} = F g_{\eta' NN} + {1 \over 2f} F^2 m_{\eta'}^2 g_{GNN}
\end{equation}
Here $F$ is a scale invariant decay constant \cite{venez} and $f=3$ is 
the number of light flavours.
The coupling $g_{\eta' NN}$ will be measured at ELSA in Bonn.

The flavour-singlet $U_A(1)$ Goldberger-Treiman relation means 
that the flavour-singlet spin structure of the nucleon 
is intimately related 
to gluodynamics and axial $U_A(1)$ symmetry.
The phenomenology of $U_A(1)$ dynamics will be explored 
in 
several new and ongoing experiments studying $\eta$ and 
$\eta'$ physics.
Photo- and leptoproduction of $\eta$ and $\eta'$ mesons 
near threshold
is being studied at ELSA \cite{elsa} and MAMI \cite{mami}. 
Higher energy measurements will be made at CEBAF \cite{cebaf}
and HERA \cite{hera}.
CELSIUS \cite{celsius} and COSY \cite{cosy} are studying
$\eta$ and $\eta'$ production in $pp$ and $pn$ scattering 
near threshold.
Central production of $\eta$ and $\eta'$ mesons in $pp$
interactions at 450 GeV/c has been measured \cite{wa102} 
by the WA102 Collaboration at CERN.
CLEO has measured the hard form-factors 
for the processes
$\eta \rightarrow \gamma \gamma^*$ 
and 
$\eta' \rightarrow \gamma \gamma^*$ 
\cite{formf}.
They have also observed strikingly large branching ratios for 
$B$ decays into an $\eta'$ and additional hadrons \cite{cleo}
\footnote{
One finds 
${\cal B}(B \rightarrow \eta' \ X)
= (6.2 \pm 1.6 \pm 1.3)$x$10^{-4}$
under the constraint
$2.0 < p_{\eta'} < 2.7$GeV/c \cite{cleo}.
Exclusive $B \rightarrow \eta' K$ 
decays have been observed
with branching ratios 
${\cal B}(B^+ \rightarrow \eta' K^+) 
  = (6.5^{+1.5}_{-1.4} \pm 0.9)$x$10^{-5}$ 
and
${\cal B}(B^0 \rightarrow \eta' K^0) 
  = (4.7^{+2.7}_{-2.0} \pm 0.9)$x$10^{-5}$.}
which may \cite{fritzschb} be related to the axial anomaly.
When combined with polarised deep inelastic scattering,
these experiments on $\eta$ and $\eta'$ production and 
decay, 
provide complementary windows on the role of gluons in dynamical 
chiral symmetry breaking.

We now review the theory of the axial anomaly and its relation
to the first moment of $g_1$.

\section{The axial anomaly}

In QCD one has to consider the effect of renormalisation.
The flavour singlet axial vector current $J_{\mu 5}^{GI}$ 
in Eqs.(10,11) 
satisfies the anomalous divergence equation 
\cite{adler,bell}
\begin{equation}
\partial^\mu J^{GI}_{\mu5}
= 2f\partial^\mu K_\mu + \sum_{i=1}^{f} 2im_i \bar{q}_i\gamma_5 q_i
\end{equation}
where
\begin{equation}
K_{\mu} = {g^2 \over 16 \pi^2}
\epsilon_{\mu \nu \rho \sigma}
\biggl[ A^{\nu}_a \biggl( \partial^{\rho} A^{\sigma}_a 
- {1 \over 3} g 
f_{abc} A^{\rho}_b A^{\sigma}_c \biggr) \biggr]
\end{equation}
is a renormalised version of the gluonic Chern-Simons
current
and the number of light flavours $f$ is $3$.
Eq.(29) allows us to write
\begin{equation}
J_{\mu 5}^{GI} = J_{\mu 5}^{\rm con} + 2f K_{\mu}
\end{equation}
where $J_{\mu 5}^{\rm con}$ and $K_{\mu}$ satisfy the
divergence equations
\begin{equation}
\partial^\mu J^{\rm con}_{\mu5}
= \sum_{i=1}^{f} 2im_i \bar{q}_i\gamma_5 q_i
\end{equation}
and
\begin{equation}
\partial^{\mu} K_{\mu} 
= {g^2 \over 8 \pi^2} G_{\mu \nu} {\tilde G}^{\mu \nu}.
\end{equation}
Here
${g^2 \over 8 \pi^2} G_{\mu \nu} {\tilde G}^{\mu \nu}$
is the topological charge density.
The partially conserved current is scale invariant 
and 
the scale dependence of $J_{\mu 5}^{GI}$ is carried entirely
by $K_{\mu}$.
When we make a gauge transformation $U$ 
the gluon field transforms as
\begin{equation}
A_{\mu} \rightarrow U A_{\mu} U^{-1} + {i \over g} (\partial_{\mu} U) U^{-1}
\end{equation}
and the operator $K_{\mu}$
transforms as
\begin{equation}
K_{\mu} \rightarrow K_{\mu} 
+ i {g \over 16 \pi^2} \epsilon_{\mu \nu \alpha \beta}
\partial^{\nu} 
\biggl( U^{\dagger} \partial^{\alpha} U A^{\beta} \biggr)
+ {1 \over 96 \pi^2} \epsilon_{\mu \nu \alpha \beta}
\biggl[ 
(U^{\dagger} \partial^{\nu} U) 
(U^{\dagger} \partial^{\alpha} U)
(U^{\dagger} \partial^{\beta} U) 
\biggr].
\end{equation}
Gauge transformations shuffle a scale invariant operator quantity
between the two operators $J_{\mu 5}^{\rm con}$ and $K_{\mu}$
whilst keeping $J_{\mu 5}^{GI}$ invariant.

The nucleon matrix element of $J_{\mu 5}^{GI}$ is
\begin{equation}
\langle p,s|J^{GI}_{5 \mu}|p',s'\rangle 
= 2m \biggl[ {\tilde s}_\mu G_A (l^2) + l_\mu l.{\tilde s} G_P (l^2) \biggr]
\end{equation}
where $l_{\mu} = (p'-p)_{\mu}$
and
${\tilde s}_{\mu} 
= {\overline u}_{(p,s)} \gamma_{\mu} \gamma_5 u_{(p',s')} / 2m $.
Since $J^{GI}_{5 \mu}$ does not couple to a massless 
Goldstone
boson it follows that $G_A(l^2)$ and $G_P(l^2)$ contain
no massless pole terms.
The forward matrix element of $J^{GI}_{5 \mu}$ is well
defined and
\begin{equation}
g_A^{(0)}|_{\rm inv} = E(\alpha_s) G_A (0).
\end{equation}

We would like to isolate the gluonic contribution to $G_A (0)$
associated with $K_{\mu}$ and thus write $g_A^{(0)}$ 
as the sum of ``quark'' and ``gluonic'' contributions.
Here one has to be careful because of the gauge dependence of
the operator $K_{\mu}$.
To understand the gluonic contributions to $g_A^{(0)}$ it is
helpful to go back to the deep inelastic cross-section in
Section 2.1.

\subsection{The anomaly and the first moment of $g_1$}

Working in the target rest frame, the spin dependent part 
of the deep inelastic cross-section, Eq.(6), is given by
\begin{equation}
{d^2 \sigma \over d\Omega dE'} 
= {\alpha^2 \over Q^4} {E' \over E} \ L_{\mu \nu}^A \ W^{\mu \nu}_A
\end{equation}
where the lepton tensor
\begin{equation}
L_{\mu \nu}^A = 2 i \epsilon_{\mu \nu \alpha \beta} k^{\alpha} q^{\beta}
\end{equation}
describes the lepton-photon vertex and the hadronic tensor
\begin{equation}
{1 \over 2m} W^{\mu \nu}_A =
i \epsilon^{\mu \nu \rho \sigma} q_{\rho} 
\biggl(
s_{\sigma} {1 \over p.q} g_1 (x,Q^2)
+ [ p.q s_{\sigma} - s.q p_{\sigma} ] {1 \over m^2 p.q} g_2 (x,Q^2) \biggr)
\end{equation}
describes the photon-nucleon interaction.

Deep inelastic scattering involves the Bjorken limit:
$Q^2 = - q^2$ and $p.q$ both $\rightarrow \infty$ 
with 
$x = {Q^2 \over 2 p.q}$ held fixed.
In terms of light-cone coordinates this corresponds to taking
$q_- \rightarrow \infty$ 
with 
$q_+ = -x p_+$ held finite.
The leading term in $W_A^{\mu \nu}$ 
is obtained by taking the Lorentz index of $s_{\sigma}$
as $\sigma = +$.
(Other terms are suppressed by powers of ${1 \over q_-}$.)

The flavour-singlet axial charge which is measured in the 
first moment of $g_1$ is 
given by the matrix element
\begin{equation}
2m s_\mu g_A^{(0)} 
= 
\langle p,s| J^{GI}_{\mu5} |p,s\rangle _c.
\nonumber
\end{equation}
If we wish to understand the first moment of $g_1$ in terms 
of the 
matrix elements of anomalous currents
($J_{\mu 5}^{\rm con}$ and $K_{\mu}$),
then we have to understand
the forward matrix element of $K_+$.

Here we are fortunate in that the parton model is formulated in the 
light-cone gauge ($A_+=0$) where the forward matrix elements of
$K_+$ are invariant.
 In the light-cone gauge the non-abelian three-gluon part of $K_+$ 
 vanishes. The forward matrix elements of $K_+$ are then invariant 
 under all residual gauge degrees of freedom.
Furthermore, in this gauge, $K_+$ measures the gluonic ``spin'' 
content
of the polarised target \cite{jafpl,man90}.
We find \cite{efremov,ccm}
\begin{equation}
G_A^{(\rm A_+ = 0)}(0) = \sum_q \Delta q_{\rm con} 
- f {\alpha_s \over 2 \pi} \Delta g
\end{equation}
where
$\Delta q_{\rm con}$ is measured by the partially conserved current
$J_{+5}^{\rm con}$
and 
$- {\alpha_s \over 2 \pi} \Delta g$ is measured by $K_+$.
The gluonic term in Eq.(42) 
offers a 
possible source for any OZI violation in $g_A^{(0)}|_{\rm inv}$.

What is the relation between the formal decomposition in Eq.(42) 
and our previous (more physical) expression (22) ?

\subsection{Questions of gauge invariance}

In perturbative QCD $\Delta q_{\rm con}$ is identified 
with 
$\Delta q_{\rm parton}$ and 
$\Delta g$ is identified with $\Delta g_{\rm parton}$ 
--- see Section 6.1 below.
If we were to work only in the light-cone gauge we might think 
that we have a complete parton model description of the first 
moment of $g_1$.
However, one is free to work in any gauge including a covariant
gauge where the forward matrix elements of $K_+$ 
are not necessarily invariant under the residual gauge degrees 
of freedom \cite{jaffem}.

We illustrate this by an example in covariant gauge.

The matrix elements of $K_{\mu}$ need to be specified with
respect to a specific gauge.
In a covariant gauge we can write
\begin{equation}
\langle p,s|K_\mu |p',s'\rangle _c
= 2m \biggl[ {\tilde s}_\mu K_A(l^2) + l_\mu l.{\tilde s} K_P(l^2) \biggr]
\end{equation}
where $K_P$ contains a massless Kogut-Susskind pole \cite{kogut}.
This massless pole cancels with a corresponding massless
pole term in $(G_P - K_P)$.
In an axial gauge $n.A=0$ the matrix elements of the gauge dependent 
operator $K_{\mu}$ will also contain terms proportional to the gauge
fixing vector $n_{\mu}$.

We may define a gauge-invariant form-factor $\chi^{g}(l^2)$
for the topological charge density (33) in the divergence of 
$K_{\mu}$:
\begin{equation}
2m l.{\tilde s} \chi^g(l^2) =
\langle p,s | {g^2 \over 8 \pi^2} G_{\mu \nu} {\tilde G}^{\mu \nu}
 | p', s' \rangle_c.
\end{equation}
Working in a covariant gauge, we find
\begin{equation}
\chi^{g}(l^2) = K_A(l^2) + l^2 K_P(l^2)
\end{equation}
by contracting Eq.(43) with $l^{\mu}$.

When we make a gauge transformation any change 
$\delta_{\rm gt}$
in $K_A(0)$ is compensated
by a corresponding change in the residue of the Kogut-Susskind
pole in $K_P$, viz.
\begin{equation}
\delta_{\rm gt} [ K_A(0) ]
+ \lim_{l^2 \rightarrow 0} \delta_{\rm gt} [ l^2 K_P(l^2) ] = 0.
\end{equation}
The Kogut-Susskind pole corresponds to the Goldstone 
boson associated with spontaneously broken $U_A(1)$ symmetry
\cite{rjc}.
There is no Kogut-Susskind pole in perturbative QCD.
It follows that the quantity which is shuffled 
between the $J_{+5}^{\rm con}$ and $K_+$ 
contributions to $g_A^{(0)}$ is strictly non-perturbative;
it vanishes in perturbative QCD and is not present in the QCD
parton model.

One can show \cite{jaffem,cron} that the forward matrix elements of 
$K_{\mu}$ are invariant under ``small'' gauge transformations
(which are topologically deformable to the identity) 
but not invariant under ``large'' gauge transformations which 
change the topological winding number.
Perturbative QCD involves only ``small'' gauge transformations;
``large'' gauge transformations involve strictly non-perturbative physics.
The second term on the right hand side of Eq.(35) is a total derivative; 
its matrix elements vanish in the forward direction.
The third term on the right hand side of Eq.(35) 
is associated with the gluon topology \cite{cron}.

The topological winding number is determined by the gluonic boundary 
conditions at 
``infinity'' 
\footnote
{A large surface with boundary which is spacelike with respect 
 to the positions $z_k$ of any operators or fields in the physical
 problem.}
\cite{Callan,rjc}.
It is insensitive to local deformations of the gluon 
field $A_{\mu}(z)$ or of the gauge transformation $U(z)$.
When we take the Fourier transform to momentum space 
the topological structure induces a light-cone zero-mode which 
can contribute to $g_1$ only at $x=0$.
Hence, we are led
to consider the possibility that there may be a 
term in $g_1$  which is proportional to $\delta(x)$ \cite{bass98}.

It remains an open question whether the net non-perturbative
quantity which
is shuffled between $K_A(0)$ and $(G_A - K_A)(0)$ under ``large''
gauge transformations
is finite or not.
If it is finite and, therefore, physical, then, when we choose
$A_+ =0$,
this non-perturbative quantity must be contained in 
some combination of the $\Delta q_{\rm con}$ and $\Delta g$ in Eq.(42).

\section{Gluon topology and $g_A^{(0)}$}

We now explain how tunneling processes may induce topological
polarisation inside a nucleon.  
This effect is related to the realisation of $U_A(1)$ symmetry 
breaking \cite{rjc,thooft,thrept} by instantons.

\subsection{$U_A(1)$ symmetry}

In classical field theory Noether's theorem tells us that there is 
a conserved current associated with each global symmetry of the 
Lagrangian.
Chiral $SU(2)_L \otimes SU(2)_R$ is associated with the isotriplet
axial vector current $J_{\mu 5}^{(3)}$.
In the classical version of QCD 
(before we turn on vacuum polarisation)
the chiral singlet $U_A(1)$ or 
$U(1)_L \otimes U(1)_R$ symmetry of ${\cal L}_{QCD}$ is associated 
with the Noether current
\begin{equation}
{\cal N}_{\mu 5} = 
\left(\bar{u}\gamma_\mu\gamma_5u
                  + \bar{d}\gamma_\mu\gamma_5d
                  + \bar{s}\gamma_\mu\gamma_5s\right)_{\rm Noether}.
\end{equation} 
This classical current satisfies the divergence equation
\begin{equation}
\partial^\mu {\cal N}_{\mu5}
= \sum_{i=1}^{f} 2im_i \bar{q}_i\gamma_5 q_i .
\end{equation}
${\cal N}_{\mu 5}$ 
is gauge invariant;
there is no anomaly in this classical theory.
The classical theory predicts the existence of 
the 
flavour-singlet, pseudoscalar Goldstone boson $\phi_0$ 
which we introduced in Section 3.4.
The axial anomaly and the absence of any such boson in 
the physical spectrum
means that the realisation of $U_A(1)$ symmetry in real 
QCD (with interactions) is quite subtle.

We choose $A_0=0$ gauge and define two operator charges:
\begin{equation}
X(t) = \int d^3z J_{05}^{GI}(z)
\end{equation}
and
\begin{equation}
Q_5(t)  = \int d^3z J_{05}^{\rm con}(z)
\end{equation}
corresponding to the gauge-invariant 
and partially conserved axial-vector currents respectively.

The charge $X(t)$ is manifestly gauge invariant
whereas $Q_5$ is only invariant under ``small'' 
gauge transformations.
It transforms as
\begin{equation}
Q_5 \rightarrow Q_5 - 2 n 
\end{equation}
where $n$ is 
the winding number associated with the gauge transformation $U$.
Whilst $Q_5$ is gauge dependent, 
we can define a gauge invariant $Q_5$ chirality, ${\cal Q}_5$, of any 
given operator ${\cal O}$ through the gauge-invariant eigenstates 
of the commutator
\begin{equation}
[Q_5 , {\cal O}]_{-} = {\cal Q}_5 \ {\cal O} .
\end{equation}
The gluon field operator and its derivative 
have zero $Q_5$ chirality and non-zero $X(t)$ chirality \cite{brandeis}.

\subsection{Instantons and $U_A(1)$ symmetry}

When topological effects are taken into account, 
the QCD vacuum $|\theta \rangle$
is a coherent superposition 
\begin{equation}
|\theta \rangle = \sum_m {\rm e}^{im \theta} |m \rangle
\end{equation}
of the
eigenstates $|m \rangle$ of 
$\int d \sigma_{\mu} K^{\mu} \neq 0$ \cite{Callan,crewther}.
Here
$\sigma_{\mu}$ is a large surface which is defined 
\cite{crewther} such that its boundary is spacelike 
with respect to the positions 
$z_k$ of
any operators or fields in the physical problem 
under discussion.
For integer values of the topological winding number $m$,
the states $|m \rangle$ contain $mf$ quark-antiquark pairs 
with non-zero $Q_5$ chirality
$\sum_l \chi_l = - 2 \xi_{\rm R} f m$ 
where $f$ is the number of light-quark flavours.
Relative to the $|m=0 \rangle$ state, the $|m=+1 \rangle$ 
state carries topological winding number +1 and $f$
quark-antiquark pairs with $Q_5$ chirality equal to $-2f \xi_{\rm R}$.
The factor $\xi_{\rm R}$ is equal to +1 if the $U_A(1)$ symmetry of 
QCD is associated with $J_{\mu 5}^{\rm con}$ and equal to -1 if the
$U_A(1)$ symmetry is associated with $J_{\mu 5}^{GI}$ --- see below.

There are two schools of thought 
\cite{rjc,thrept}
about how instantons break $U_A(1)$ symmetry.
Both of these schools start from 't Hooft's 
observation \cite{thooft} 
that the flavour determinant
\begin{equation}
\langle {\rm det} \biggl[
{\rm {\overline q}_L}^i {\rm q_R}^j {\rm (z)} 
\biggr]
\rangle_{\rm inst.} \neq 0
\end{equation}
in the presence of a vacuum tunneling process 
between states with different topological winding number.
(We denote the tunneling process by the subscript ``inst.''.
 It is not specified at this stage whether ``inst.'' denotes 
 an instanton or an anti-instanton.)

\begin{enumerate}
\item
{\bf Explicit $U_A(1)$ symmetry breaking}

In this scenario \cite{thooft,thrept} the $U_A(1)$ symmetry of QCD 
is associated with the current $J_{\mu 5}^{GI}$ and the topological 
charge density is treated like a mass term in the divergence of 
$J_{\mu 5}^{GI}$.
The quark chiralities which appear in the flavour 
determinant (54) are associated with $X(t)$
so that the net axial charge $g_A^{(0)}$ is not conserved 
$(\Delta X \neq 0)$
and the net $Q_5$ chirality is conserved $(\Delta Q_5 = 0)$ 
in quark instanton scattering processes.

In QCD with $f$ light flavoured quarks the (anti-)instanton 
``vertex''  involves a total of $2f$ light quarks and 
antiquarks. 
Consider a flavour-singlet combination of $f$ right-handed 
($Q_5 =+1$) quarks incident on an anti-instanton.
The final state for this process consists of a
flavour-singlet combination of
$f$ left-handed ($Q_5 = -1$) quarks;
$+2f$ units of $Q_5$ chirality are taken away by an effective 
``schizon'' 
\cite{thrept}
which carries zero energy and zero momentum.
The ``schizon'' is introduced to ensure $Q_5$ conservation.
Energy and momentum are conserved between the in-state and
out-state quarks in the quark-instanton scattering process.
The non-conservation of $g_A^{(0)}$ is ensured by a term coupled 
to $K_{\mu}$ with
equal magnitude and opposite sign to the ``schizon'' 
term which 
also carries zero energy and zero momentum.
This gluonic term describes the change in the topological
winding number which is induced by the tunneling process.
The anti-instanton changes the net $U_A(1)$ chirality by 
an amount $(\Delta X = -2f)$.

This picture is the basis of 't Hooft's effective instanton
interaction \cite{thooft}.

\item
{\bf Spontaneous $U_A(1)$ symmetry breaking}

In this scenario the $U_A(1)$ symmetry of QCD is associated with the 
partially-conserved axial-vector current $J_{\mu 5}^{\rm con}$. 
Here, the quark chiralities 
which appear in the flavour determinant (54) are identified with $Q_5$. 
With this identification,
the net axial charge $g_A^{(0)}$ is conserved 
$(\Delta X = 0)$
and the
net $Q_5$ chirality is not conserved $(\Delta Q_5 \neq 0)$ 
in 
quark instanton
scattering processes.
This result is the opposite to what happens in the explicit
symmetry breaking scenario.
When $f$ right-handed quarks scatter on an instanton 
\footnote{cf. an anti-instanton in the explicit $U_A(1)$ 
 symmetry breaking scenario.}
the final state involves $f$ left-handed quarks.
There is no ``schizon'' and the instanton induces a change 
in the net $Q_5$ 
chirality
$\Delta Q_5 = -2f$.
The conservation of $g_A^{(0)}$ is ensured by the gluonic term 
coupled to $K_{\mu}$ 
which measures the change in the topological winding number
and which carries zero energy and zero momentum.
The charge $Q_5$ is time independent for massless quarks 
(where $J_{\mu 5}^{\rm con}$ is conserved).
Since $\Delta Q_5 \neq 0$ in quark instanton scattering 
processes we find that the $U_A(1)$ symmetry is spontaneously 
broken by instantons.
The Goldstone boson is manifest \cite{rjc} as the massless 
Kogut-Susskind pole which 
couples to $J_{\mu 5}^{\rm con}$ and $K_{\mu}$ but not to $J_{\mu 5}^{GI}$
--- see Eq.(43).

\end{enumerate}

It is important to note that the $X(t)$ and $Q_5$ chiralities have different 
physical content. The difference between the two theories of quark instanton 
interactions is about more than just the sign of the ($X$ or $Q_5$) chirality 
which is flipped in the quark instanton scattering process.
We now explain why these two possible realisations of $U_A(1)$ 
symmetry breaking have a different signature in $\nu p$ elastic scattering.

\subsection{The topological contribution to $g_A^{(0)}$}

In both the explicit and spontaneous symmetry breaking scenarios
we may consider multiple scattering of the incident quark first 
from an instanton and then from an anti-instanton.
Let this process recur a large number of times.
When we time-average over a large number of such
interactions, then the time averaged expectation
value of the chirality $Q_5$
carried by the incident quark is reduced from the naive value $+1$
that it would take in the absence of vacuum tunneling processes.
Indeed, in one flavour QCD the time averaged value of $Q_5$ tends 
to zero at large times \cite{forte1,forte2}.

In the spontaneous $U_A(1)$ symmetry breaking scenario \cite{rjc}
any instanton induced suppression of the flavour-singlet axial 
charge which is measured in polarised deep inelastic scattering 
is compensated by a net transfer of axial charge or ``spin'' from 
partons carrying finite momentum fraction $x$ to the flavour-singlet
topological
term at $x=0$.
It induces a flavour-singlet $\delta (x)$ term in $g_1$ which is not 
present in 
the explicit $U_A(1)$ symmetry breaking scenario.

The net topological term is gauge invariant.
In the $A_0=0$ gauge the $x=0$ 
polarisation 
is ``gluonic'' 
and is measured by $\int d^3z K_0$.
In the light-cone gauge this polarisation 
may be re-distributed between the ``quark'' and ``gluonic'' 
terms measured by $J_{+ 5}^{\rm con}$ and $K_+$ respectively.

To guide our intuition about non-perturbative QCD it is sometimes helpful 
to consider analogies with similar phenomena in condensed matter physics.
For example, the Nambu-Jona-Lasino (NJL) model \cite{njla,njlb,njlc} 
is motivated by the analogy between the constituent quarks and the 
Dirac quasi-particles which appear in the BCS theory of superconductivity
\cite{bcs}.
Keeping in mind that the underlying physics is fundamentally different,
it is nevertheless interesting to note that 
polarised zero-momentum modes are observed in low temperature physics 
in the form of polarised condensates. 
The vacuum of the A-phase of superfluid $^3$He behaves both as an orbital 
ferromagnet and uniaxial liquid crystal with spontaneous magnetisation 
along the anistropy axis ${\hat l}$, and as a spin antiferromagnet with 
magnetic anisotropy along a second axis ${\hat d}$ \cite{anderson}.
Recent experiments \cite{he4} have revealed that superfluidity 
in $^4$He can form in finite systems; 60 atoms of $^4$He are the minimum 
needed for superfluidity.

\subsection{How to measure topological polarisation}

The scale-invariant flavour singlet axial charge can be measured
independently in 
elastic Z$^0$ exchange processes such as elastic neutrino proton 
scattering 
\cite{garvey} and parity violation in light atoms \cite{parity,nachtmann}.
QCD renormalisation group arguments tell us that
the neutral current axial charge which is measured 
in elastic
$\nu p$ scattering is \cite{kaplan}
\begin{eqnarray}
g_A^{(Z)} 
&=& {1 \over 2} \ g_A^{(3)} \ + \ {1 \over 6} \ g_A^{(8)} \
- \ {1 \over 6} \ (1 + {\rm C}) \ g_A^{(0)}|_{\rm inv}  \  + \ 
{\cal O}({1 \over m_h}) \\
\nonumber
&=& {1 \over 2} \ \biggl( g_A^{(3)} - \Delta s|_{\rm inv} \biggr)
  \ - {1 \over 6} \ {\rm C} \ g_A^{(0)}|_{\rm inv}  
  \  + {\cal O}({1 \over m_h}) .
\end{eqnarray}
Here ${\rm C}$ denotes the leading order heavy-quark 
contributions 
to $g_A^{(Z)}$ and $m_h$ represents the heavy-quark mass.
Numerically,
${\rm C}$ is a $\simeq 6-10\%$ correction \cite{kaplan}
---
within the present experimental error on $g_A^{(0)}|_{\rm inv}$.
The flavour-singlet axial charge in Eq.(55) includes
any contribution
from the topological term at $x=0$ 
\footnote{
Heavy-quark instanton interactions are suppressed as 
${\cal O}(1/m_h)$ where $m_h$ is the heavy-quark mass \cite{svz}.
It follows that the coefficient of
any heavy-quark $\delta(x)$
term in $g_1$ decouples as ${\cal O}(1/m_h)$.
It does not affect the relation between polarised
deep inelastic scattering and $\nu p$ elastic scattering.}
.
(In $\nu p$ elastic scattering there is no kinematic factor which 
 could filter out zero mode contributions to $g_A^{(0)}$,  unlike 
 deep inelastic scattering where Bjorken $x=0$ is kinematically 
 unreachable at finite $Q^2$.)

If the topological contribution ${\cal C}$ to $g_A^{(0)}$ 
is finite,
then
the flavour-singlet axial charge
which is 
extracted from a polarised deep inelastic experiment is
$(g_A^{(0)} - {\cal C})$.
Elastic ${\rm Z}^0$ exchange processes such as $\nu p$ 
elastic scattering \cite{garvey} 
and parity violation in light atoms \cite{parity,nachtmann} 
measure the full $g_A^{(0)}$ \cite{kaplan}.
One can measure the 
effect of the topological $x=0$ polarisation
by comparing the flavour-singlet axial charges
which are extracted from polarised deep inelastic and
$\nu p$ elastic scattering experiments.

\section{Partons and $g_1$}

\subsection{The QCD parton model}

The parton model description of polarised deep inelastic scattering
involves writing the deep inelastic structure functions as the sum
over the convolution of ``soft'' quark and gluon parton distributions
with ``hard'' photon-parton scattering coefficients.
We focus on the flavour-singlet part of $g_1$
\begin{equation}
g_1|_{\rm singlet} = {1 \over 9}
\biggl( \sum_q \Delta q \otimes C^q + N_f \Delta g \otimes C^g
\biggr)
.
\end{equation}
Here, $\Delta q(x)$ and $\Delta g(x)$ denote the quark and gluon parton
distributions, $C^q$ and $C^g$ denote the corresponding hard
scattering coefficients, and $N_f$ is the number of quark flavours
liberated into the final state.
The parton distributions are target dependent and describe a flux of
quark and gluon partons into the hard (target independent)
photon-parton interaction which is described by the coefficients.
The separation of $g_1$ into ``hard'' and ``soft'' is not unique
and
depends on the choice of factorisation scheme \cite{bint,mank,bodwin,manoh}.

The hard coefficients are calculable in perturbative QCD.
One can use a kinematic cut-off on the partons' transverse momentum  
squared ($k_T^2 > \lambda^2$) 
to define the factorisation scheme and thus separate the hard 
and 
soft parts of the phase space for the photon-parton collision. 
The cut-off $\lambda^2$ is called the factorisation scale.
The coefficients have the perturbative expansion
$C^q = \delta(1-x) +
         {\alpha_s \over 2\pi} f^q(x, {Q^2 / \lambda^2})$ 
and 
  $C^g = {\alpha_s \over 2\pi} f^g(x, {Q^2 / \lambda^2})$
where
the functions $f^q$ and $f^g$ have at 
most a $\ln (1-x)$ singularity when $x \rightarrow 1$ \cite{ratc}.

The gluon coefficient is calculated from the box graph for photon-gluon 
fusion. 
We use a cut-off on the transverse momentum squared of the struck quark 
relative to the photon-gluon direction to separate the total phase space 
into ``hard'' ($k_T^2 \geq \lambda^2$) and ``soft'' ($k_T^2 < \lambda^2$) 
contributions.
One finds \cite{bnt}:
\begin{eqnarray}
g_1^{(\gamma^* g)} (x,Q^{2},P^{2})|_{\rm hard} &=&
-{\alpha_s \over 2 \pi }
{\cut \over \cxx} \Biggl[ (2x-1)(\cx) \\ \nonumber
& &
\biggl(1 - {1 \over {\cut \sqrt{\cxx} }}
\ln \biggl({ {1+\sqrt{\cxx} \cut}\over {1-\sqrt{\cxx} \cut}}
\biggr) \biggr) \\ \nonumber
& &
+ (x-1+{{x P^{2}}\over{Q^{2}}})
{{\left( 2m^{2}(\cxx)- P^{2}x(2x-1)(\cx)\right)}
\over {(m^{2} + \lambda^2) (\cxx) - P^{2}x(x-1+{{x P^{2}}\over{Q^{2}}})}}
\Biggr]
\end{eqnarray}
for each flavour of quark liberated into the final state.
Here
$m$ is the quark mass, 
$P^2 = -p^2$ is the virtuality of the gluon,
$x$ is the Bjorken variable ($x= {Q^2 \over 2 \nu}$)
and
$s$ is the centre of mass energy squared
$s= (p+q)^2 = Q^2 ( {1 - x \over x} ) - P^2$
for the photon-gluon collision.

If we set $\lambda^2$ to zero, thus including the entire phase space,
then we obtain the full box graph contribution to $g_1^{\gamma^* g}$.
The gluon structure function $g_1^{(\gamma^* g)}$ is invariant under
the exchange of
($p \leftrightarrow q$).
If we take $\lambda^2$ to be finite and independent of $x$,
then the crossing symmetry of
$g_1^{(\gamma^* g)}$ under the exchange of ($p \leftrightarrow q$)
is realised separately in each of the ``hard'' and ``soft'' parts 
of $g_1^{(\gamma^* g)}$.

When $Q^2$ is much greater than the other scales ($\lambda^2, m^2, P^2$)
in Eq.(57)
the expression for $g_1^{(\gamma^* g)}$ simplifies to
\begin{eqnarray}
g_1^{(\gamma^* g)}|_{\rm hard} &=&
{\alpha_s \over 2 \pi} 
\Biggl[ 
  (2x-1) \Biggl( \ln {Q^2 \over \lambda^2} + \ln {1-x \over x} - 1 \Biggr)
\\ \nonumber
&+& (2x-1) \ln {\lambda^2 \over {x(1-x) P^2 + (m^2 + \lambda^2)} }
 +  (1 -x) { {2m^2 - P^2x(2x-1)} \over {x(1-x)P^2 + m^2 + \lambda^2} }
\Biggr] .
\end{eqnarray}
We choose an $x$-independent cut-off ($Q^2 \gg \lambda^2, m^2, P^2$).
The first moment of $g_1^{(\gamma^* g)}|_{\rm hard}$ 
is the sum of two contributions \cite{bint}:
\begin{equation}
\int_0^1 dx g_1^{(\gamma^{*} g)}|_{\rm hard} =
- {\alpha_s \over 2 \pi}
\left[1 +
\frac{2m^{2}}{P^{2}}
\frac{1}{\sqrt{1+4(m^{2}+\lambda^2)/P^{2}}}
\ln \left(
\frac{1 - \sqrt{1+4(m^{2}+\lambda^2)/P^{2}}}
{1+\sqrt{1+4(m^{2}+\lambda^2)/P^{2}}}
\right)\right].
\end{equation}
The unity term describes a contact photon-gluon interaction.
It comes from the region of phase space where the hard photon 
scatters on a quark or antiquark carrying transverse momentum 
squared $k_T^2 \sim Q^2$ \cite{ccm}.
The second term comes from the kinematic region 
$k_T^2 \sim {\cal O}(\lambda^2, P^2,m^2)$.
It vanishes when we take the factorisation scale 
$\lambda^2 \gg P^2, m^2$.
The $- {\alpha_s \over 2 \pi}$ 
factor in Eq.(59) 
is the coefficient of $\Delta g_{\rm parton}$ in Eq.(22) and 
$\Delta g$ in Eq.(42).

When we apply the operator product expansion the first term in 
Eq.(59)
corresponds to the gluon matrix element of the anomaly current 
$K_{\mu}$.
If we remove the cut-off by setting $\lambda^2$ equal to zero, 
then the second term in Eq.(59) is the gluon matrix element of 
$J_{\mu 5}^{\rm con}$ \cite{bint}.
This term is associated with the ``soft'' quark distribution of 
the gluon
$\Delta q^{(g)}(x, \lambda^2)$.
By extending this operator product expansion analysis to the 
higher moments of 
$g_1^{\gamma^{*}g}$,
one can show that \cite{bass92,cheng,chengx}
that the axial anomaly contribution to the {\it shape} of $g_1$ 
at finite $x$ is given by the convolution of the polarised gluon 
distribution $\Delta g(x,Q^2)$ with the hard coefficient
\begin{equation}
{\tilde C}^{(g)}|_{\rm anom} = - {\alpha_s \over \pi} (1-x) .
\end{equation}
This anomaly contribution is a small $x$ effect in $g_1$; 
it is
essentially negligible for $x$ less than 0.05
\cite{bnt,strat,grsv,eks}.
The hard coefficient ${\tilde C}^{(g)}|_{\rm anom}$ 
is normally included as a term in $C^g$ --- Eq.(56).
It is associated with two-quark jet events carrying $k_T^2 \sim Q^2$
in the final state.

One could also consider an $x$-dependent cut-off on the struck 
quark's virtuality
\cite{bint,bnt}
\begin{equation}
m^2 - k^2 
= P^2 x + {k_T^2 + m^2 \over (1-x)} > \lambda_0^2 
= {\rm constant}(x)
\end{equation}
or a cut-off on the invariant mass squared 
of the quark-antiquark
component of 
the light-cone wavefunction of the target gluon \cite{mank,lepage}
\begin{equation}
{\cal M}^2_{q {\overline q}} 
= {k_T^2 + m^2 \over x (1-x)} + P^2 \geq \lambda_0^2 
= {\rm constant}(x) .
\end{equation}
These different choices of infrared cut-offs correspond to different
jet definitions and different factorisation schemes for photon-gluon 
fusion.
If we evaluate the first moment of $g_1^{(\gamma^{*} g)}$ 
using the cut-off on the quarks' virtuality,
then we find
``half of the anomaly'' 
in the gluon coefficient through the mixing 
of transverse and longitudinal momentum components \cite{bint,bnt}.
The anomaly coefficient for the first moment is recovered with the
invariant mass squared cut-off through a sensitive cancellation of
large and
small $x$ contributions \cite{bint}.

The $x$-independent cut-off is preferred for discussions of the
axial anomaly and the symmetry properties of the $\gamma^* g$ 
interaction.
The reason for this is that the transverse momentum is defined
perpendicular to the plane spanned by $p_{\mu}$ and $q_{\mu}$
in momentum space.
The $x$-dependent cut-offs mix the transverse and longitudinal
components of momentum.
Substituting Eqs.(61,62) into Eq.(57) we find that the ``hard''
and ``soft'' contributions to $g_1^{(\gamma^* g)}$ do not separately
satisfy the
$(p \leftrightarrow q)$ symmetry of $g_1 (x,Q^{2})$ if use the
$x$-dependent cut-offs to define the ``hard'' part of the total
phase space \cite{bassbs}.

\subsection{QCD evolution}

In deep inelastic scattering experiments the different $x$ data 
points on $g_1$ are each measured at different values of $Q^2$, 
viz. $x_{\rm expt.}(Q^2)$.
One has to evolve these experimental data points 
to the same value of $Q^2$ 
in order to test the Bjorken \cite{bj} and Ellis-Jaffe \cite{ej} sum-rules.

The structure function $g_1$ is given by the sum of the convolution 
of the parton distributions $\Delta q$ and $\Delta g$
with the hard scattering coefficients $C^q$ and $C^g$
respectively --- see Eq.(56).
The structure function is dependent on $Q^2$ and independent of 
the factorisation scale $\lambda^2$ and the ``scheme'' used to
separate
the $\gamma^{*}$-parton cross-section into ``hard'' and ``soft''
contributions.
Examples of different ``schemes'' are the transverse momentum
squared and virtuality cut-offs that we discussed in Section 6.1.

In the parton model formula (56) the hard coefficients $C^q$ and 
$C^g$ are calculable in perturbative QCD as a function of $Q^2$ 
and the factorisation scale $\lambda^2$.
The $\lambda^2$ dependence of the parton distributions is given 
by the DGLAP equations \cite{altp}
\begin{eqnarray}
{d \over dt} \Delta \Sigma(x,t) &=& {\alpha_s(t) \over 2 \pi}
\biggl[
\int_x^1 {dy \over y} \Delta P_{qq}({x \over y}) \Delta \Sigma (y,t) 
+ 2 N_f 
\int_x^1 {dy \over y} \Delta P_{qg}({x \over y}) \Delta g (y,t)
\biggr]
\\ \nonumber
{d \over dt} \Delta g (x,t) &=& {\alpha_s(t) \over 2 \pi}
\biggl[
\int_x^1 {dy \over y}
\Delta P_{gq}({x \over y}) \Delta \Sigma (y,t) 
+ 
\int_x^1 {dy \over y} \Delta P_{gg}({x \over y}) \Delta g (y,t) \biggr]
\end{eqnarray}
where $\Sigma(x,t) = \sum_q \Delta q(x,t)$ and $t = \ln \lambda^2$.
The splitting functions $P_{ij}$ in Eq.(63) have been calculated at
next-to-leading order by Mertig, Zijlstra and van Neervan \cite{neervan} 
and by Vogelsang \cite{vogelsang}.

\subsection{Gluonic contributions to $g_1$}

The size of $\Delta g_{\rm parton}$ is one of the key issues in QCD 
spin physics at the present time.

The polarised gluon distribution $\Delta g(x,Q\lambda^2)$ contributes 
to $g_1$ 
through the convolution $\Delta g \otimes C^g$.
Depending on the choice of factorisation 
scheme,
the gluonic coefficient $C^g$ has at most a $\ln (1-x)$ 
singularity when $x \rightarrow 1$.
In contrast,
the leading term in the quark coefficient $C^q$ is $\delta (1-x)$.
The convolution involving $C^g$ has the practical effect that
$\Delta g$
makes a direct
contribution to $g_1$ only at $x < 0.05$ \cite{bnt,strat,grsv,eks}.

At the same time, the $\lambda^2$ evolution of the flavour-singlet quark 
distribution involves the polarised gluon distribution 
$\Delta g(y,\lambda^2)$ at values of $y$ in the range ($x<y<1$) \cite{altp}
--- see Eq.(63).
Thus, through evolution, the polarised gluon distribution is relevant 
to the shape of $g_1$ 
over the complete $x$ range.
This result enables one to carry out next-to-leading order QCD 
fits to polarised deep inelastic data with the hope of extracting 
some information about $\Delta g_{\rm parton}$.
Here, one starts with an ansatz for the shape of $\Delta q(x,Q_0^2)$ 
and $\Delta g(x,Q_0^2)$ at some particular input scale $Q_0^2$.
The input distributions are evolved to the range of $Q^2$ covered 
by the deep inelastic experiments.
Finally, one chooses a particular factorisation scheme
(see below) and makes a best fit to the $g_1$ data in terms of 
the input shape parameters and the scale $Q_0^2$.

Several groups have followed this approach [4,104,105,112-117].
Different QCD motivated fits to the polarised deep inelastic data 
yield values of 
$\Delta g_{\rm parton}(Q^2)$  
between zero and +2 at $Q^2 = 1$GeV$^2$.
The value of $\Delta g_{\rm parton}$ which is extracted from these 
fits depends strongly on the functional form which is assumed 
for the input distributions with only small changes in the overall 
$\chi^2$ for the fits \cite{defl}
--- we refer to De Florian et al. \cite{defl} 
for a nice overview of QCD fits to $g_1$ data.

Three schemes are commonly used in the analysis of experimental 
data: the $k_T^2$ cut-off, ${\overline {\rm MS}}$ and AB schemes.
These schemes correspond to different procedures for
separating the phase space for photon-gluon fusion into ``hard''
and ``soft'' contributions.

In the parton model that we discussed in Section 6.1 using the 
cut-off on the transverse momentum squared, 
the polarised gluon contribution to the first moment of $g_1$ 
is associated with two-quark jet events carrying $k_T^2 \sim Q^2$.
The hard coefficient is given by
$C^{(g)}_{{\rm PM}} = g_1^{(\gamma^{*} g)}|_{\rm hard}$ 
in Eq.(58) with $Q^2 \geq \lambda^2$ and $\lambda^2 \gg P^2, m^2$.
This ``parton model scheme'' is sometimes called 
the ``chiral invariant'' (CI) \cite{cheng} or JET \cite{leader} scheme.

Different schemes can be defined relative to this ``parton model scheme'' 
by the transformation
\begin{equation}
C^{(g)} \biggl( x, {Q^2 \over \lambda^2}, \alpha_s(\lambda^2) \biggr) 
\rightarrow 
C^{(g)} \biggl( x, {Q^2 \over \lambda^2}, \alpha_s(\lambda^2) \biggr) 
- 
{\tilde C}^{(g)}_{\rm scheme} \biggl( x, \alpha_s(\lambda^2) \biggr)
\end{equation}
where ${\tilde C}^{(g)}_{\rm scheme}$ 
equals ${\alpha_s \over \pi}$ times a polynomial in $x$.
The parton distributions transform under (64) as
\begin{eqnarray}
\Delta \Sigma (x,\lambda^2)_{\rm scheme} 
&=& \Delta \Sigma (x,\lambda^2)_{\rm PM} + 
  N_f 
  \int_x^1 {dz \over z} \Delta g({x \over z},\lambda^2)_{\rm PM} 
  {\tilde C}^{(g)}_{\rm scheme} (z, \alpha_s(\lambda^2)) 
\\ \nonumber
\Delta g (x,\lambda^2)_{\rm scheme} &=& \Delta g(x,\lambda^2)_{\rm PM}
\end{eqnarray}
so that $g_1$ is left invariant by the change of scheme.
The virtuality and invariant-mass cut-off versions of the
parton model that we discussed in Section 6.1 correspond to 
different 
choices of scheme.

The ${\overline {\rm MS}}$ and AB schemes are defined as follows.
In the ${\overline {\rm MS}}$ scheme the gluonic hard scattering 
coefficient is calculated using 
the operator product expansion with ${\overline {\rm MS}}$ 
renormalisation \cite{thv}. 
One finds \cite{cheng,bass92}:
\begin{equation}
C^{(g)}_{{\overline{\rm MS}}} 
=
C^{(g)}_{\rm PM}
\ + \ 
{\alpha_s \over \pi} (1-x) .
\end{equation}
In this scheme
$\int_0^1 dx \ C^{(g)}_{\overline {\rm MS}} =0$ 
so that
$\int_0^1 dx \ \Delta g(x, \lambda^2)$ decouples from $\int_0^1 dx g_1$.
This result corresponds to the fact that there is no gauge-invariant 
twist-two, spin-one, gluonic operator with $J^P = 1^+$ 
to appear in 
the operator product expansion for the first moment of $g_1$.
In the ${\overline {\rm MS}}$ scheme the 
contribution of $\int_0^1 dx \ \Delta g$
to the first moment of $g_1$ is included 
into 
$\int_0^1 dx \ \Sigma_{{\overline{\rm MS}}} (x,\lambda^2)$. 
The AB scheme \cite{forteab} is defined by the formal operation of
adding the $x$-independent term $-{\alpha_s \over 2 \pi}$ to the 
${\overline {\rm MS}}$ gluonic coefficient,
viz.
\begin{equation}
C^{(g)}_{\rm AB}(x) = 
C^{(g)}_{{\overline{\rm MS}}}
\ - \ {\alpha_s \over 2 \pi}.
\end{equation}
In both the parton model and AB schemes 
$\int_0^1 dx \ C^{(g)} = - {\alpha_s \over 2 \pi}$.
We refer to Cheng \cite{chengx} and Llewellyn Smith \cite{llew}
for a critical discussion of these schemes and their application
to polarised deep inelastic scattering.

The SMC and SLAC E-154 experiments quote values of $\Delta g$ 
for their own data.
The SMC values are \cite{smcqcd}
\begin{equation}
\Delta g_{\rm parton} 
= + \ 0.25 \ ^{+0.29}_{-0.22} \ \ \ \ , \ \ \ \  
{\overline{\rm MS} \ {\rm scheme}}
\end{equation}
and
\begin{equation}
\Delta g_{\rm parton} 
= + \ 0.99 \ ^{+1.17}_{-0.70} \ \ \ \ , \ \ \ \ {\rm AB \ scheme}
\end{equation}
in the ${\overline{\rm MS}}$ \cite{thv} and AB \cite{forteab} schemes 
respectively --- 
each at 1GeV$^2$:
The E-154 values are \cite{e154qcd}
\begin{equation}
\Delta g_{\rm parton} 
= + 1.8 \ ^{+0.7}_{-1.0} \ \ \ \ , \ \ \ \  {\overline{\rm MS} \ {\rm scheme}}
\end{equation}
and
\begin{equation}
\Delta g_{\rm parton} 
= + 0.4 \ ^{+1.7}_{-0.9} \ \ \ \ , \ \ \ \ {\rm AB \ scheme}
\end{equation}
--- each at 5GeV$^2$.

Dedicated experiments have been proposed to measure $\Delta g_{\rm parton}$
more precisely.
The COMPASS \cite{compass} and HERMES \cite{hermesc} 
experiments will measure charm production in polarised 
deep inelastic scattering;
a further experiment is proposed for SLAC \cite{bosted}.
Polarised RHIC \cite{rhic} 
will measure prompt photon production in polarised $pp$
collisions through 
the process $q g \rightarrow q \gamma$,
thus enabling a different measurement of $\Delta g_{\rm parton}$.
In the longer term there is a proposal to polarise the proton beam at 
HERA \cite{albert,heraproc}.
A polarised proton beam at HERA would allow precision measurements of
$g_1$ at small $x$ where it becomes increasingly more sensitive to 
the polarised gluon distribution and to study the two-quark-jet 
cross-section associated with the axial anomaly.

\section{The shape of $g_1$}

There have been many theoretical papers proposing possible explanations
of the small value of $g_A^{(0)}|_{\rm pDIS}$ extracted from polarised 
deep inelastic scattering.
Besides offering an explanation of the size of $g_A^{(0)}$ it is 
important to understand how the different possible effects 
contribute to the shape of $g_1$, which contains considerably 
more information than just the first moment.

Early constituent quark model calculations of the shape of $g_1$ 
\cite{kutiw,closev,carlitz}
still provide a reasonable description of 
the $g_1$ data in the ``valence'' region ($x$ greater than about 0.2).
More recent quark model calculations 
\cite{bagsfa,reinhardt}
include QCD 
evolution from the Bag 
``input scale'' $\mu_0^2$ to deep inelastic $Q^2$.
In their next-to-leading order Bag model fits to deep inelastic data 
Steffens et al. \cite{bagsfa} 
found that the model ``input scale'' increased corresponding 
to a change in the coupling $\alpha_s(\mu_0^2)$ from 0.8 to 0.6 
when pion cloud corrections were included into the model input.

Semi-inclusive measurements of $g_1$ will enable us to disentangle
the separate 
$\Delta q_{\rm valence}(x)$ and $\Delta {\overline q}_{\rm sea}(x)$
contributions to $g_1$ \cite{semia,semib}.
The first semi-inclusive measurements have been published by the SMC 
\cite{semiexpt}; more precise data will soon be available from HERMES.

If the sum of 
$\Delta u_{\rm v} = \int_0^1 dx \Delta u_{\rm valence}(x)$ 
and
$\Delta d_{\rm v} = \int_0^1 dx \Delta d_{\rm valence}(x)$ 
extracted from semi-inclusive measurements of $g_1$ 
falls short of 
the constituent quark model prediction for $g_A^{(0)}$,
then the ``discrepancy'' could be interpreted as a 
first hint that some of the nucleon's spin might reside 
at $x=0$.
(Recall from Section 5 that instanton tunneling processes
 have the potential to shift some fraction of $g_A^{(0)}$ 
 from valence partons carrying $x>0$ to the topological 
 term at $x=0$.)
The first semi-inclusive data (from SMC) 
yield
$\Delta u_{\rm v} = +0.77 \pm 0.10 \pm 0.08$ 
and
$\Delta d_{\rm v} = -0.52 \pm 0.14 \pm 0.09$.
Combining the errors in quadrature we find
$\Delta u_{\rm v} + \Delta d_{v} = +0.25 \pm 0.21$ 
(cf. the constituent quark model prediction $g_A^{(0)} \simeq 0.6$)
and
$\Delta u_{\rm v} - \Delta d_{v} = +1.29 \pm 0.21$
(cf. $g_A^{(3)} = 1.267 \pm 0.004$).
It will be interesting to see how these results hold up in the light 
of more accurate data from HERMES.

Wislicki \cite{wislicki} has analysed the polarised semi-inclusive 
data from
SMC and HERMES looking for possible evidence of the axial anomaly 
at large $x$.
The data shows no evidence of any deviation between the charge parity 
$C=+1$ and $C=-1$ polarised quark distributions in the ``valence'' 
region $x > 0.3$.

Perturbative QCD Counting Rules \cite{brodbs} make predictions for 
the large $x$ behaviour of $g_1$.
The small $x$ extrapolation of $g_1$ data is presently the largest
source of experimental error on deep inelastic measurements of the 
nucleon's axial charges.
The small $x$ extrapolation is usually motivated either by 
Regge theory 
or
by perturbative QCD arguments.

We now outline what is known about $g_1$ at large $x$ (Section 7.1) 
and at small $x$ (Section 7.2).
In Section 7.3 we collect this theory and describe how it explains 
the shape of the measured spin dependent and spin 
independent isotriplet structure functions as a function of $x$.

\subsection{$g_1$ at $x \rightarrow 1$}

Perturbative QCD counting rules predict that the parton distributions 
should 
behave as a power series expansion in $(1-x)$ when $x \rightarrow 1$
\cite{brodbs}.
We use $q^{\uparrow}(x)$ and $q^{\downarrow}(x)$ to denote the parton 
distributions
polarised parallel and antiparallel to the polarised proton.
One finds \cite{brodbs}
\begin{equation}
q^{\uparrow \downarrow} (x) \rightarrow (1-x)^{2n - 1 + \Delta S_z}
\ \ \ \ ; \ \ \ \ (x \rightarrow 1).
\end{equation}
Here, $n$ is the number of spectators and $\Delta S_z$ is the difference 
between the polarisation of the struck quark and the polarisation of the 
target nucleon.
When $x \rightarrow 1$ the QCD counting rules predict that the structure 
functions should be dominated by valence quarks polarised parallel to the 
spin of the nucleon.
The ratio of polarised to unpolarised structure functions should go to one
when $x \rightarrow 1$.

\subsection{$g_1$ at small $x$}

The small $x$ extrapolation of $g_1$ data is important for precise 
measurements of the nucleon's axial charges from deep inelastic 
scattering.
The SLAC data \cite{e143a,e154} has the smallest experimental error 
in the $x$ range $(0.01 < x < 0.12)$.
We show these data in Fig.1.

There are several important properties of the $g_1$ data at small $x$.
\begin{enumerate}
\item
The magnitude of $g_1^{(p-n)}$ is significantly 
greater 
than 
the magnitude of
$g_1^{(p+n)}$ in the measured small $x$ region.
This is in contrast to unpolarised deep inelastic scattering
where the small $x$ region is dominated by isoscalar pomeron 
exchange.
\item
The isosinglet $g_1^{(p+n)}$ is small and consistent with 
zero in the measured small $x$ range ($0.01 < x < 0.05$).
Polarised gluon models \cite{abfr}
predict that $g_1^{(p+n)}$ may become strongly 
negative at smaller values of $x$ ($\sim 10^{-4}$)
but this remains to be checked experimentally.
\item
Polarised deep inelastic data from CERN and SLAC consistently 
indicate a strong isotriplet term in $g_1$ which rises at small $x$.
\end{enumerate}

We consider the isotriplet part of $g_1$ in more detail.
Combining the proton data from E-143 \cite{e143a}
together with the 
neutron 
data from E-154 \cite{e154}, 
one finds a good fit \cite{soff,mart} to the SLAC data on 
$g_1^{(p-n)}$:
\begin{equation}
g_1^{(p-n)} \sim (0.14) \ x^{- {1 \over 2}}
\end{equation}
in the $x$ range $(0.01 < x < 0.12)$ at $Q^2 \simeq 5$GeV$^2$.

Regge theory makes a prediction for the large $s_{\gamma p}$ 
dependence of the spin dependent and spin independent parts 
of the total photoproduction $(Q^2=0)$ cross-section.
It is often used to describe the small $x$ behaviour of deep 
inelastic structure functions ($Q^2$ larger than about 2GeV$^2$).
The Regge prediction for the isotriplet part of $g_1$ is
\cite{heim,ek}
\begin{equation}
g_1^{(p-n)} \sim x^{- \alpha_{a_1}} \ \ \ \ , \ \ \ \ (x \rightarrow 0) .
\end{equation}
Here $\alpha_{a_1}$ is the intercept of the $a_1$ 
Regge trajectory.
If one makes the usual assumption that the $a_1$ trajectory 
is a straight line running parallel to the $(\rho, \omega)$ 
trajectories,
then one finds $\alpha_{a_1} = -0.4$.

Clearly, Regge theory does not provide a good description of $g_1^{(p-n)}$
in the measured $x$ range  $(0.01 < x < 0.12)$.
At first glance, this result is surprising since Regge theory provides 
a good description \cite{pvl1} of the NMC measurements \cite{nmcf2} of 
both the isotriplet and isosinglet parts of $F_2$ in the same small $x$ 
range $(0.01 < x < 0.1)$ at $Q^2 \simeq 5$GeV$^2$.
In practice, the shape of $g_1$ at small $x$ is $Q^2$ dependent.
The $Q^2$ dependence is driven by
DGLAP evolution and, 
at very small $x$ 
($\sim 10^{-3}$), 
by the resummation of $\alpha_s^{l} \ln^{2l} x$ radiative corrections 
--- see eg. \cite{grsv,badk,zeut}.

To understand this evolution, let us define an effective intercept 
${\tilde \alpha_{a_1}}(Q^2)$ 
to describe the small $x$ behaviour of $g_1$ at finite $Q^2$:
$g_1^{(p-n)} \sim x^{- {\tilde \alpha_{a_1}}}$. 
The net $Q^2$ dependence of ${\tilde \alpha_{a_1}}$ depends strongly 
on the 
value of
${\tilde \alpha_{a_1}}$ 
which is needed to describe the leading twist part of $g_1^{(p-n)}$ 
at low momentum scales --- 
for example $\mu_0^2 \sim 0.3$GeV$^2$.
Let $(\Delta u - \Delta d)(x)$ denote the 
leading twist (=2) part of $g_1^{(p-n)}$.
DGLAP evolution of $(\Delta u - \Delta d)(x)$ from $\mu_0^2$ to deep 
inelastic $Q^2$ shifts the weight of the distribution from larger to 
smaller values of $x$ whilst keeping the area under the curve,
$g_A^{(3)}$, 
constant.
QCD evolution has the practical effect of ``filling up'' the small $x$
region --- increasing the value of ${\tilde \alpha_{a_1}}$ 
with increasing $Q^2$.
The scale independence of $g_A^{(3)}$ provides an important constraint 
on the change in 
${\tilde \alpha_{a_1}}$ under QCD evolution.
The closer that ${\tilde \alpha_{a_1}}(\mu_0^2)$ 
is to the Regge prediction -0.4, 
the more that ${\tilde \alpha_{a_1}}(Q^2)$
will grow in order to preserve the area under 
$(\Delta u - \Delta d)(x)$ 
when we increase $Q^2$ to values typical of deep inelastic scattering.

Badelek and Kwiecinski \cite{badk} have investigated the effect of 
DGLAP and $\alpha_s \ln^2x$ resummation on the small $x$ behaviour 
of $g_1^{(p-n)}$.
They find a good fit to the data using a flat small-$x$ input 
distribution 
at 
$Q_0^2 = 1$GeV$^2$.
In their optimal NLO QCD fit to polarised deep inelastic data 
Gl\"uck et al.
\cite{grsv} used a rising input at $\mu_0^2 \simeq 0.3$GeV$^2$.

The isosinglet part of $g_1$ is more complicated because of possible
gluonic exchanges in the $t-$channel.
There have been several suggestions how the isosinglet part of 
$g_1$ should behave at small $x$ based on non-perturbative 
\cite{kuti,clos,sbpl} and perturbative \cite{grsv,kirs,bart,blum} 
QCD arguments.

\subsection{Isotriplet structure functions}

To understand the shape of $g_1^{(p-n)}$ it is helpful to compare the
isotriplet part of $g_1$ 
with the 
isotriplet part of $F_2$ (the nucleon's spin independent structure function).

In the QCD parton model
\begin{equation}
2x (g_1^p - g_1^n) = 
{1 \over 3} x 
\Biggl[ (u + {\overline u})^{\uparrow} - (u + {\overline u})^{\downarrow} 
      - (d + {\overline d})^{\uparrow} + (d + {\overline
         d})^{\downarrow}
 \Biggr] \otimes \Delta C_{NS}
\end{equation}
and
\begin{equation}
(F_2^p - F_2^n) 
= 
{1 \over 3} x 
\Biggl[ (u + {\overline u})^{\uparrow} + (u + {\overline u})^{\downarrow} 
      - (d + {\overline d})^{\uparrow} - (d + {\overline
         d})^{\downarrow}
 \Biggr] \otimes C_{NS}.
\end{equation}
Here $u$ and $d$ denote the up and down flavoured quark 
distributions
polarised parallel $({\uparrow})$ and antiparallel $({\downarrow})$
to the target proton
and 
$\Delta C_{NS}$ and $C_{NS}$ 
denote the spin-dependent and 
spin-independent perturbative QCD coefficients \cite{ratc}
\footnote{
The coefficients 
$C_{NS}$ and $\Delta C_{NS}$ have the perturbative expansion
$\delta (1-x) + {\alpha_s \over 2 \pi} f(x)$.
They 
are related 
(in the ${\overline{\rm MS}}$ scheme)
by \cite{ratc}
$
\Delta C_{NS} (x) = C_{NS}(x) - {\alpha_s \over 2 \pi}
{4 \over 3}(1 + x).
$
The difference between $C_{NS}$ and $\Delta C_{NS}$ makes a
non-negligible contribution to the deep inelastic structure 
functions only at $x < 0.05$.
}
.
There is no gluonic or pomeron contribution to the isotriplet 
structure functions $g_1^{(p-n)}$ and $F_2^{(p-n)}$.

In Fig.2 we show the SLAC data \cite{e143a,e154} on $g_1^{(p-n)}(x)$ 
together 
with the NMC measurement \cite{nmc} of $F_2^{(p-n)}(x)$. 
The NMC data are quoted at $Q^2=4$GeV$^2$.
Our SLAC data set is obtained by combining the published E-143 data 
on $g_1^p$ with the E-154 data on $g_1^n$ --- both at $Q^2=5$GeV$^2$.
We combine the $g_1$ measurements to produce 
one $g_1^{(p-n)}$
data point for each $x$ bin listed by the NMC.
Clearly, $2xg_1^{(p-n)}$ is greater than $F_2^{(p-n)}$ for $x < 0.4$
in this data 
(--- see also \cite{mart,smcfinal}).

Sum-rules for the first moments of $g_1^{(p-n)}$ and the unpolarised
structure function $F_2^{(p-n)}/x$
provide important constraints for our understanding of the structure
of the nucleon.
The Gottfried integral \cite{gott}
\begin{eqnarray}
I_G 
&=& \int_0^1 dx \Biggl( {F_2^p - F_2^n \over x} \Biggr) \\ \nonumber
&=& {1 \over 3} \int_0^1 dx \Biggl( u_V (x) - d_V (x) \Biggr)
  + {2 \over 3} \int_0^1 dx 
    \Biggl( {\overline u} (x) - {\overline d} (x) \Biggr)  
\end{eqnarray}
measures any SU(2) flavour asymmetry in the sea.
The integral $I_G$ has been measured by the NMC in deep inelastic 
scattering 
($I_G =  0.235 \pm 0.026$) \cite{nmc}
and by the Fermilab E-866 NuSea Collaboration in Drell-Yan
production
($I_G =  0.267 \pm 0.018$) \cite{nusea}.
Possible explanations of this effect include the pion cloud of 
the nucleon \cite{pion} and the Pauli-principle \cite{feynman}
at work in the nucleon's 
sea.
We refer to Thomas and Melnitchouk \cite{shimoda} for a pedagogical 
review of the physics involved in understanding the Gottfried integral.

When we convolute the polarised and unpolarised parton distributions
with the 
perturbative coefficients $C_{NS}$ and $\Delta C_{NS}$ 
the difference between the two coefficients makes a negligible 
contribution to the structure functions at $x > 0.05$ --- the bulk of 
the $x$ range in Fig.2.
Dividing ${1 \over 3} g_A^{(3)}$ by the central values of 
$I_G$ measured by the NMC and Fermilab E-866 experiments 
we obtain 1.78 and 1.57 respectively.
Whilst the physical values of $g_A^{(3)}$ and $I_G$ differ markedly
from the simple SU(6) predictions, it is interesting to observe that 
the ratio ${1 \over 3} g_A^{(3)}/I_G$ is consistent with the SU(6) 
prediction ($2I_{Bj}/I_G = {5 \over 3}$).

In Fig.3 we plot the ratio $R_{(3)}(x) = 2 x g_1^{(p-n)} / F_2^{(p-n)}$.
We also plot the 
SU(6) prediction for $2 I_{Bj} / I_G$, 
together with the result of dividing twice the value of 
${1 \over 6}g_A^{(3)}$ by the central values of the measured Gottfried 
integral $I_G$ from NMC 
(top line) 
and E-866 (bottom line).

There are several observations to make.
First,
the deep inelastic data is consistent 
with 
$R_{(3)}(x) \simeq {5 \over 3}$ in the $x$ range $(0.02 < x < 0.4)$
\cite{mart}.
At large $x$
the data is consistent with the QCD Counting Rules prediction
\begin{equation}
R_{3} \rightarrow 1 \ \ \ \ , \ \ \ \ {x \rightarrow 1}.
\end{equation}
The data exhibits no evidence of the simple Regge prediction
$R_{(3)} \propto x$ when $x \rightarrow 0$ (--- say $x<0.1$).
The SMC have recently published their measurements of $R_{(3)}$ 
down to $x=0.005$.
This new data is consistent with the SLAC measurements and the
observation 
$R_{(3)} \simeq {5 \over 3}$.

We make some phenomenological observations which may help 
to understand the relative shapes of the isotriplet parts 
of $g_1$ and $F_2$.
First, the total area under $g_1^{(p-n)}$ is fixed by the 
Bjorken sum-rule.
Soffer and Teryaev \cite{soff} have observed that $\simeq 50\%$ 
of
the Bjorken sum-rule comes from small $x$ ($x < 0.12$) if the
shape (68) is extrapolated to $x=0$.
Suppose that we pivot $g_1^{(p-n)}$ about its measured value at 
$x=0.12$
and impose the 
Regge behaviour $\sim x^{+0.4}$ at smaller values of $x$ 
instead of the observed small $x$ behaviour $\sim x^{- 0.5}$.
We shall call this the ``Regge modified'' $g_1^{(p-n)}$.
For the ``Regge modified'' $g_1^{(p-n)}$ the fraction of the 
total Bjorken sum-rule which comes from $x$ less than 0.12 
would be reduced to $\simeq 17\%$.
That is, $\simeq 33\%$ of the total Bjorken sum-rule would be 
shifted to larger values of $x$.

A priori, one would expect the simple SU(6) prediction for $R_{(3)}$ 
to come closest to the ratio of the leading twist contribution to the 
measured structure functions at a value $x^*$ close to ${1 \over 3}$ 
{\it after} the leading twist parton distributions have been evolved
to the quark model scale $\mu_0^2$. 
QCD evolution shifts the value of $x^*$ to slightly smaller $x$ at deep 
inelastic $Q^2$.
Deep inelastic structure functions fall rapidly to zero when $x \rightarrow 1$.
Suppose we combine the ``Regge modified'' 
$g_1^{(p-n)}$ 
together with the measured $F_2^{(p-n)}$ data.
The resultant ``Regge modified'' $R_{(3)}$ would considerably exceed 
the SU(6) prediction $R_{(3)} = {5 \over 3}$
in the intermediate $x$ region because of the extra area 
that has been shifted under $g_1^{(p-n)}$ at $x > 0.12$.
However, this contains the $x$ range where we would most expect the SU(6) 
prediction to work 
(if it is to work at all).
In summary,
{\it if} $R_{(3)}$ takes the SU(6) value ${5 \over 3}$ at some 
intermediate value $x^*$ and 
decreases towards one when we increase $x$ greater than $x^*$,
and if
$g_1^{(p-n)}$ has a power law behaviour $\sim x^{- {\tilde \alpha}_{a_1}}$
at small $x$,
then $g_1^{(p-n)}$ must rise at small $x$ 
(in contradiction to Regge theory) 
with $R_{(3)} \sim {5 \over 3}$
so that $g_1^{(p-n)}$ saturates the Bjorken sum-rule with the physical value 
of $g_A^{(3)}$.

To summarise this Section, constituent quark model calculations 
provide a reasonable description of $g_1$ in the intermediate $x$ 
region.
The soft Regge theory predictions for the small $x$ behaviour of 
the $g_1$ spin structure function seem to fail badly at deep 
inelastic $Q^2$.
The shape of the measured isotriplet spin structure function
$g_1^{(p-n)}$ 
may be understood in terms of perturbative QCD Counting Rules 
(at $x>0.2$) and constituent quark model ideas (in the $x$ range 
 $0.01 < x < 0.2$).

\section{Conclusions and Outlook}

Relativistic constituent-quark pion coupling models 
predict
$g_A^{(0)}|_{\rm inv} \simeq g_A^{(8)} \simeq 0.6$.
The value of $g_A^{(0)}$ extracted from polarised
deep inelastic scattering experiments is 
$g_A^{(0)}|_{\rm pDIS} \simeq 0.2$ -- 0.35.
This result has inspired many theoretical ideas about the internal
spin structure of the nucleon.
Central to these ideas is the role that the axial anomaly plays in 
the transition from parton to constituent quark degrees of freedom 
in low energy QCD.

In QCD some fraction of the spin of the nucleon and of the 
constituent quark may be carried by gluon topology.
If the topological contribution ${\cal C}$ is indeed finite, 
then the constituent quark model predictions for $g_A^{(0)}$
are not necessarily in contradiction with the small value of
$g_A^{(0)}|_{\rm pDIS} = (g_A^{(0)} - {\cal C})$
extracted from polarised deep inelastic scattering.

New experiments will help to further resolve the spin structure
of the nucleon and to distinguish between the various theoretical
possibilities.
\begin{itemize}
\item
Semi-inclusive measurements of $g_1$ in the current fragmentation 
region (HERMES) will enable more accurate measurements of the 
valence and sea quark contributions to $g_1$.
If a polarised proton beam becomes available at HERA it will be
possible to extend this programme into the target fragmentation
region and to study the target (in-)dependence of the small value
of $g_A^{(0)}|_{\rm pDIS}$.
\item
Measurements of open charm production in polarised deep inelastic
scattering (COMPASS, HERMES and SLAC) will enable a direct 
measurement of the polarised gluon distribution $\Delta g(x)$ in
deep inelastic scattering.
A complementary measurement of $\Delta g(x)$ will come from studies 
of prompt photon production in polarised $pp$ collisions at RHIC.
It will be interesting to compare the deep inelastic and polarised
$pp$ measurements of $\Delta g_{\rm parton}$.
A polarised proton beam at HERA would enable us to measure $g_1$ 
at small $x$ where it is most sensitive to $\Delta g(x)$ and 
to study the two-quark-jet cross-section associated with the axial anomaly.
\item
A precision measurement of $\nu p$ elastic scattering or parity 
violation in light atoms would enable us to make an independent 
determination of 
$g_A^{(0)}|_{\rm inv}$.
The value of $g_A^{(0)}|_{\rm inv}$ which is extracted from these 
elastic Z$^0$ exchange processes includes any contribution from 
topological polarisation at $x=0$
whereas the deep inelastic measurement does not 
--- thus, enabling us to measure the topological term ${\cal C}$.
\end{itemize}
The physics of the flavour-singlet axial-charge $g_A^{(0)}$ provides a 
bridge between the internal spin structure of the nucleon and chiral 
$U_A(1)$ dynamics. When combined with experimental and theoretical 
studies of the $\eta - \eta'$ system, QCD spin physics offers a 
new window on the role of gluons in dynamical chiral symmetry breaking.

\vspace{1.0cm}

{\bf Acknowledgements:} \\

I thank B. Povh for the invitation to write this review and for his 
hospitality at the MPI, Heidelberg.
Many of the ideas in this review have been developed in fruitful
collaboration with 
M.M. Brisudova, S.J. Brodsky, R.J. Crewther, A. De Roeck, P.V. Landshoff, 
I. Schmidt, F.M. Steffens and A.W. Thomas. 
I also thank 
V.N. Gribov, H. Fritzsch, P. Minkowski, B. Povh, D. Sch\"utte, G.M. Shore 
and W. Weise
for helpful discussions about the proton spin problem in QCD and 
constituent quark models, and N. Bianchi, A. Br\"ull, G. Garvey,
M. Moinester, J. Pochodzalla and R. Windmolders for discussions 
about experimental data.

\newpage

\begin{figure}[tbh] 
\includegraphics{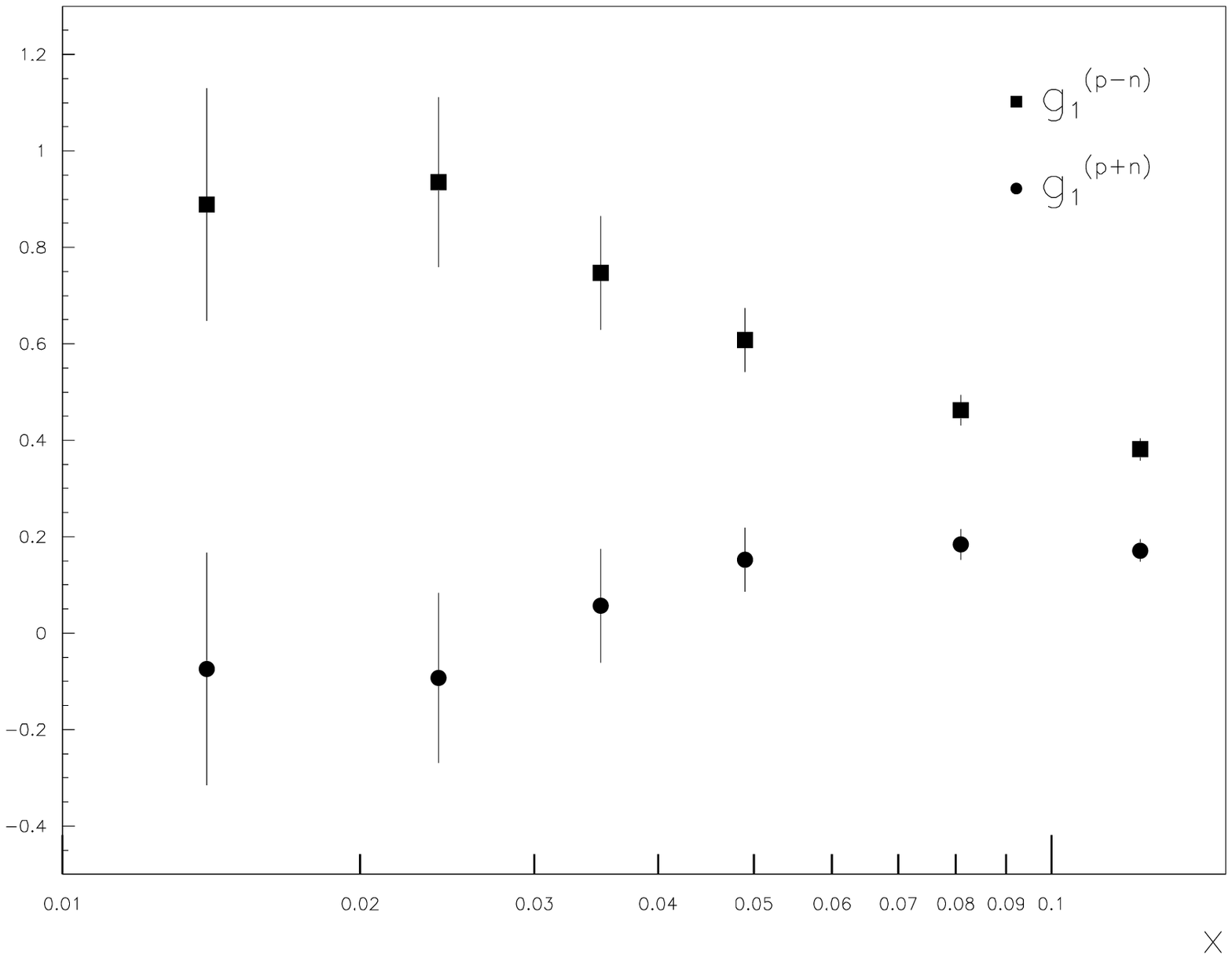} 
\begin{center} 
\vspace{10cm} 
\parbox{15cm} 
{\caption[Delta]
{
The SLAC data on $g_1^{(p-n)}$ at small $x$.
The proton data is taken from E-143 \cite{e143a}
and E-155 \cite{e155} (two smallest $x$ data points).
The neutron data is taken from E-154 \cite{e154}.
}
\label{fig1}} 
\end{center} 
\end{figure}

\newpage

\begin{figure}[tbh] 
\includegraphics{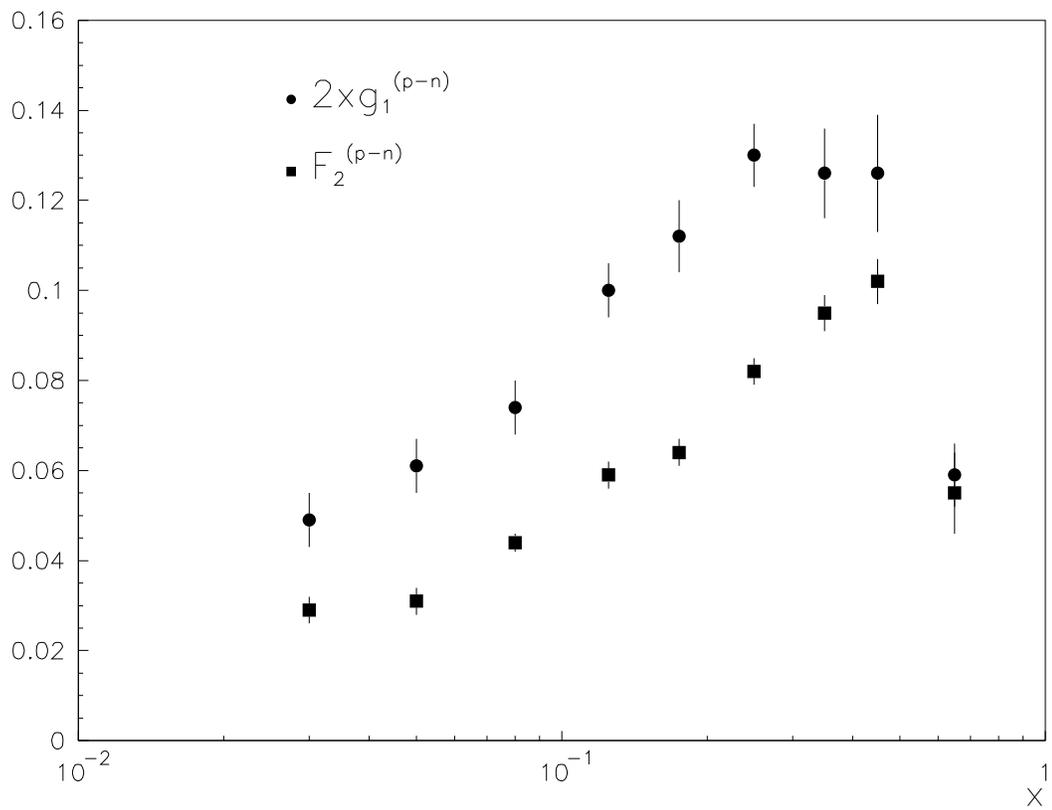} 
\begin{center} 
\vspace{10cm} 
\parbox{15cm} 
{\caption[Delta]
{
The isotriplet structure functions $2x g_1^{(p-n)}$ and $F_2^{(p-n)}$.
The $g_1^{(p-n)}$ data is from SLAC, the $F_2^{(p-n)}$ data is from NMC.
}
\label{fig2}} 
\end{center} 
\end{figure}

\newpage

\begin{figure}[tbh] 
\includegraphics{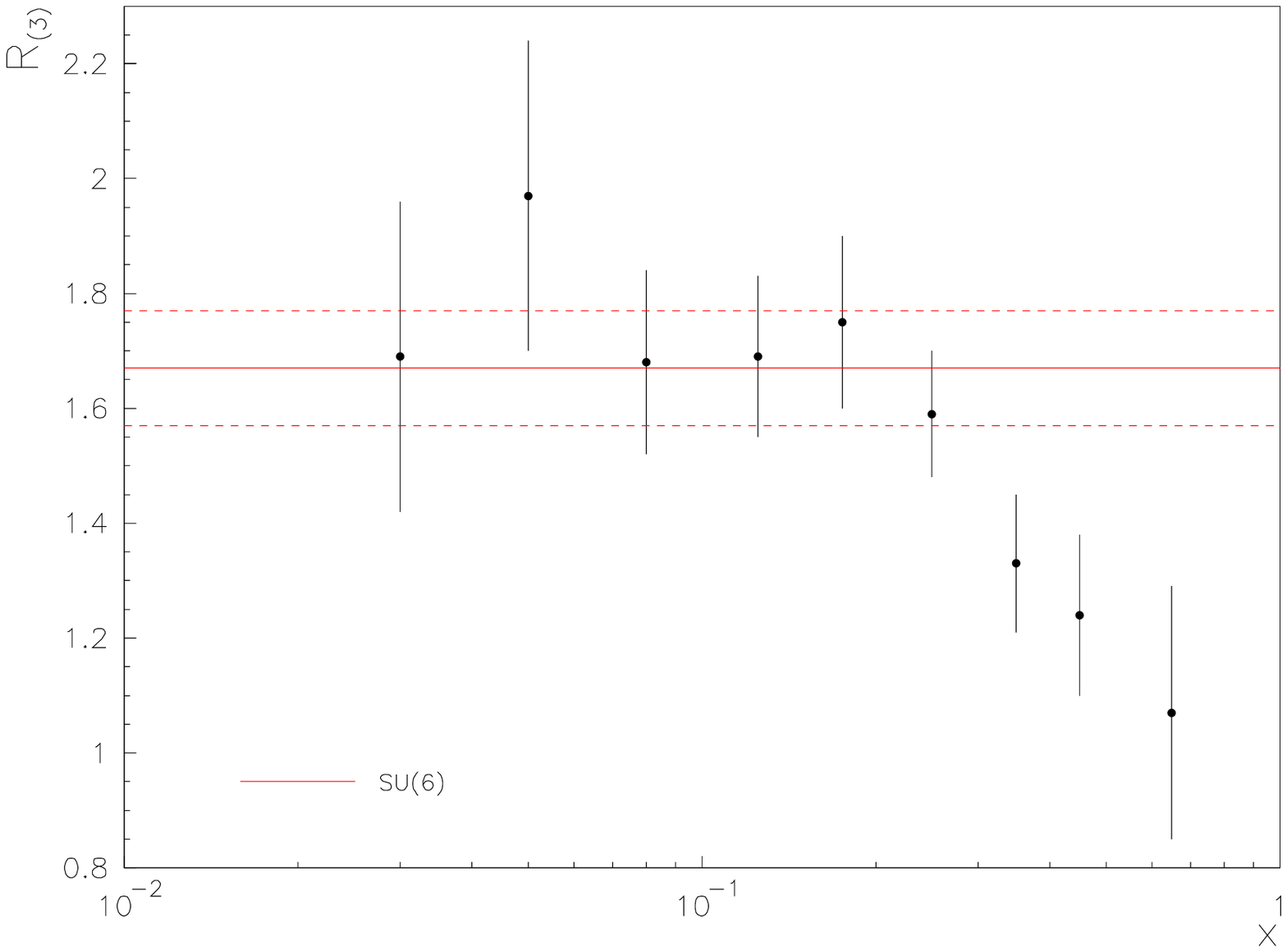}
\begin{center} 
\vspace{10cm} 
\parbox{15cm} 
{\caption[Delta]
{ 
The ratio $R_{(3)} = 2x g_1^{(p-n)}/F_2^{(p-n)}$ obtained 
from the $g_1^{(p-n)}$ and $F_2^{(p-n)}$ data in Fig.2.
}
\label{fig3}} 
\end{center} 
\end{figure}


\newpage


\begin{thebibliography}{99}
%
\bibitem{emc} EMC Collaboration
(J Ashman et al.) Phys. Lett. {\bf B206} (1988) 364;
Nucl. Phys. {\bf B328} (1989) 1.
%
\bibitem{smc}
The Spin Muon Collaboration (D. Adams et al.),
Phys. Lett. {\bf B396} (1997) 338; \\
(B. Adeva et al.), Phys. Lett. {\bf B412} (1997) 414.
%
\bibitem{smcfinal}
The Spin Muon Collaboration (B. Adeva et al), 
Phys. Rev. {\bf D58} (1998) 112001
%
\bibitem{smcqcd}
The Spin Muon Collaboration (B. Adeva et al.), 
Phys. Rev. {\bf D58} (1998) 112002.
%
\bibitem{hermes}
The HERMES Collaboration (K. Ackerstaff et al.),
Phys. Lett. {\bf B404} (1997) 383; \\
The HERMES Collaboration (A. Airapetian et al.),
Phys. Lett. {\bf B442} (1998) 484.
%
\bibitem{baum}
The E-130 Collaboration (G. Baum et al.),
Phys. Rev. Lett. {\bf 51} (1983) 1135.
%
\bibitem{e142}
The E-142 Collaboration (P.L. Anthony et al.),
Phys. Rev. {\bf D54} (1996) 6620.
%
\bibitem{e143a}
The E-143 Collaboration (K. Abe et al.),
Phys. Rev. Lett. {\bf 74} (1995) 346; \\
Phys. Rev. {\bf D58} (1998) 112003.
%
\bibitem{e154}
The E-154 Collaboration (K. Abe et al.),
Phys. Rev. Lett. {\bf 79} (1997) 26.
%
\bibitem{partal}
G. Altarelli, Phys. Rep. {\bf 81} (1982) 1.
%
\bibitem{cloudy}
A.W. Thomas, Adv. Nucl. Phys. {\bf 13} (1984) 1.
%
\bibitem{njlb}
U. Vogl and W. Weise, Prog. Part. Nucl. Phys. {\bf 27} (1991) 195.
%
\bibitem{njlc}
S.P. Klevansky, Rev. Mod. Phys. {\bf 64} (1992) 649.
%
\bibitem{Callan}
C.G. Callan, R.F. Dashen and D.J. Gross, Phys. Lett. {\bf B63} (1976) 334; \\
R. Jackiw and C. Rebbi, Phys. Rev. Lett. {\bf 37} (1976) 172.
%
\bibitem{rjc}
R.J. Crewther, Effects of Topological Charge in Gauge
Theories, in Facts and Prospects of Gauge Theories, Schladming,
Austria, February 1978, ed. P. Urban, Acta Physica Austriaca Suppl.
{\bf 19} (1978) 47.
%
\bibitem{njla}
Y. Nambu and G. Jona-Lasinio, Phys. Rev. {\bf 122} (1961) 345.
%
\bibitem{adler} S.L. Adler, Phys. Rev. {\bf 177} (1969) 2426.
%
\bibitem{bell} J.S. Bell and R. Jackiw, Nuovo Cimento {\bf 60A} (1969)
47.
%
\bibitem{bass98} 
S.D. Bass, Mod. Phys. Lett. {\bf A13} (1998) 791.
%
\bibitem{schreiber}
A.W. Schreiber and A.W. Thomas, Phys. Lett. {\bf B215} (1988) 141.
%
\bibitem{weise}
K. Steininger and W. Weise, Phys. Rev. {\bf D48} (1993) 1433; \\
K. Suzuki and W. Weise, Nucl. Phys. {\bf A634} (1998) 141.
%
\bibitem{efremov}
A.V. Efremov and O.V. Teryaev, JINR Report
E2--88--287 (1988), and in Proceedings of the International Hadron
Symposium, Bechyn\v{e} 1988, eds.\ J. Fischer et al.\ 
(Czechoslovakian Academy of Science, Prague, 1989) p. 302; \\
A.V. Efremov, J. Soffer and O.V. Teryaev, Nucl. Phys. {\bf 346} (1990) 97.
%
\bibitem{ar}
G. Altarelli and G.G. Ross, Phys. Lett. {\bf B212} (1988) 391.
%
\bibitem{ccm}
R.D. Carlitz, J.C. Collins, and A.H. Mueller, Phys. Lett. {\bf B214}
(1988) 229.
%
\bibitem{jaffem} 
R.L. Jaffe and A. Manohar, Nucl. Phys. {\bf B337} (1990) 509.
%
\bibitem{venez}
G. Veneziano, Mod. Phys. Lett. {\bf A4} (1989) 1605; \\
G.M. Shore and G. Veneziano, Nucl. Phys. {\bf B381} (1992) 23.
%
\bibitem{narison}
S. Narison, G.M. Shore and G. Veneziano, Nucl. Phys. {\bf B433}  (1995) 209.
%
\bibitem{fritzsch}
H. Fritzsch, Phys. Lett. {\bf B229} (1989) 122, 
               {\it ibid} {\bf B256} (1991) 75.
%
\bibitem{forte1}
S. Forte, Phys. Lett. {\bf B224} (1989) 189;
          Nucl. Phys. {\bf B331} (1990) 1.
%
\bibitem{forte2}
S. Forte and E.V. Shuryak, Nucl. Phys. {\bf B357} (1991) 153.
%
\bibitem{ioffe}
B.L. Ioffe and M. Karliner, Phys. Lett. {\bf B247} (1990) 387.
%
\bibitem{altr}
G. Altarelli and G. Ridolfi, Nucl. Phys. B (Proc. Suppl.) 
{\bf 39B,C} (1995) 106.
%
\bibitem{cheng}
H.-Y. Cheng, Int. J. Mod. Phys. {\bf A11} (1996) 5109.
%
\bibitem{anselmino}
M. Anselmino, A. Efremov and E. Leader, Phys. Rept. {\bf 261} (1995) 1.
%
\bibitem{jaffe}
R.L. Jaffe, hep-ph/9602236,
{\it in} Proc. Erice School {\it The spin structure of the nucleon}, 
eds. B. Frois and V. Hughes (World Scientific, 1997).
%
\bibitem{ellisk}
J. Ellis and M. Karliner, hep-ph/9601280,
{\it in} Proc. Erice School {\it The spin structure of the nucleon}, 
eds. B. Frois and V. Hughes (World Scientific, 1997).
%
\bibitem{reyal}
B. Lampe and E. Reya, hep-ph/9810270.
%
\bibitem{shorerev}
G.M. Shore,
Zuoz lecture hep-ph/9812354 and Erice lecture hep-ph/9812355.
%
\bibitem{bassdhg}
S.D. Bass, Mod. Phys. Lett. {\bf A12} (1997) 1051.
%
\bibitem{roberts}
R.G. Roberts, ``The structure of the proton'' (Cambridge UP, 1990)
%
\bibitem{deroeck}
A.M. Cooper-Sarker, R.C.E. Devenish and A. De Roeck,
Int J. Mod. Phys. {\bf A13} (1998) 3385.
%
\bibitem{exptg2}
The E-143 Collaboration (K. Abe et al.), Phys. Rev. Lett. {\bf 76} (1996) 587; \\
The E-154 Collaboration (K. Abe et al.), Phys. Lett. {\bf B404} (1997) 377; \\
The SMC Collaboration (B. Adeva et al.), Phys. Lett. {\bf B336} (1994 125.  
%
\bibitem{jaffeg2}
R.L. Jaffe, Comm. Nucl. Part. Phys. {\bf 19} (1990) 239.
%
\bibitem{bj} 
J.D. Bjorken, Phys. Rev. {\bf 148} (1966) 1467; 
Phys. Rev. {\bf D1} (1970) 1376.
%
\bibitem{ej} J. Ellis and R.L. Jaffe, Phys. Rev. {\bf D9} (1974) 1444;
(E) {\bf D10} (1974) 1669.
%
\bibitem{kod} J. Kodaira, Nucl. Phys. {\bf B165} (1980) 129.
%
\bibitem{larin} S.A. Larin, Phys. Lett. {\bf B334} (1994) 192;
{\it ibid} {\bf B404} (1997) 153.
%
\bibitem{datagroup}
The Particle Data Group (C. Caso et al.), Euro. Phys. J. {\bf C3} (1998) 1.
%
\bibitem{su3}
F.E. Close and R.G. Roberts, Phys. Lett. {\bf B316} (1993) 165.
%
\bibitem{minka} P. Minkowski, {\em in\/} Proc. Workshop on {\em
Effective Field Theories of the Standard Model\/}, Dobog\'{o}k\~{o},
Hungary 1991, ed. U.-G. Meissner (World Scientific, Singapore, 1992).
%
\bibitem{koeb}
R. K\"{o}berle and N.K. Nielsen, Phys. Rev. {\bf D8}
(1973) 660; \\
R.J. Crewther, S.-S. Shei and T.-M. Yan, Phys. Rev. {\bf D8} (1973) 3396;\\
R.J. Crewther and N.K. Nielsen, Nucl. Phys. {\bf B87} (1975) 52.
%
\bibitem{kaplan}
D.B. Kaplan and A.V. Manohar, Nucl. Phys. {\bf B310} (1988) 527.
%
\bibitem{rjcthanks}
I thank R.J. Crewther for emphasising this point.
%
\bibitem{schlumpf}
S.J. Brodsky and F. Schlumpf, Phys. Lett. {\bf B329} (1994) 111.
%
\bibitem{henley}
W. Koepf, E.M. Henley and S.J. Pollock, Phys. Lett. {\bf B288} (1992) 11.
%
\bibitem{grv}
M. Gl\"uck, E. Reya and A. Vogt, Z Physik {\bf C67} (1995) 433.
%
\bibitem{bagsf}
R.L. Jaffe and G.G. Ross, Phys. Lett. {\bf B93} (1980) 313; \\
A.W. Schreiber, A.I. Signal and A.W. Thomas, Phys. Rev {\bf D44} (1991) 2653.
%
\bibitem{bagsfa}
F.M. Steffens, H. Holtmann and A.W. Thomas, Phys. Lett.{\bf B358} (1995) 139.
%
\bibitem{gribov}
V.N. Gribov, Physica Scripta {\bf T15} (1987) 164; \\
Lund preprint LU-TP-91-7 (1991) unpublished.
%
\bibitem{garvey}
G.T. Garvey, W.C. Louis and D.H. White, Phys. Rev. {\bf C48} (1993) 761.
%
\bibitem{parity}
B.A. Campbell, J. Ellis and R.A. Flores, Phys. Lett. {\bf B225} (1989) 419.
%
\bibitem{nachtmann}
D. Bruss, T. Gasenzer and O. Nachtmann,
Phys. Lett. {\bf A239} (1998) 81; hep-ph/9802317.
%
\bibitem{al}
G. Altarelli and B. Lampe, Z. Phys. {\bf C47} (1990) 315; \\
 G. Altarelli, Proc. Physics at HERA, Vol. 1 (1991) 379, eds.
 W. Buchm\"uller and G. Ingelman.
%
\bibitem{broadhurst}
G.C. Fox and D.Z. Freedman, Phys. Rev. {\bf 182} (1969) 1628; \\
D. Broadhurst, J. Gunion and R.L. Jaffe, Phys. Rev. {\bf D8} (1973) 566;
Ann. Phys. {\bf 81} (1973) 88.
%
\bibitem{dashen}
S.L. Adler and R.F. Dashen, {\it Current Algebras and Applications
to Particle Physics} (W.A. Benjamin, 1968).
%
\bibitem{fmink}
H. Fritzsch and P. Minkowski, Nuovo Cim. {\bf 30A} (1975) 393.
%
\bibitem{christos}
G. Christos, Phys. Rept. {\bf 116} (1984) 251.
%
\bibitem{witten}
E. Witten, Nucl. Phys. {\bf B156} (1979) 269; \\
G. Veneziano, Nucl. Phys. {\bf B159} (1979) 213.
%
\bibitem{thooft}
G. 't Hooft, Phys. Rev. Lett. {\bf 37} (1976) 8; 
             Phys. Rev. {\bf D14} (1976) 3432.
%
\bibitem{thrept}
G. 't Hooft, Phys. Rept. {\bf 142} (1986) 357.
%
\bibitem{mink}
P. Minkowski, Phys. Lett. {\bf B237} (1990) 531; {\bf B423} (1998) 157.
%
\bibitem{bek}
S.J. Brodsky, J. Ellis and M. Karliner, Phys. Lett. {\bf B206} (1988) 309.
%
\bibitem{gsven}
G.M. Shore and G. Veneziano, Nucl. Phys. {\bf B516} (1998) 333.
%
\bibitem{ozidep}
D. de Florian, G.M. Shore and G. Veneziano, hep-ph/9711353,
{\it in} Proc. Workshop on 
{\it Physics with Polarized Protons at HERA}, 
eds. A. De Roeck and T. Gehrmann, DESY-Proceedings-1998-01
%
\bibitem{albert}
A. De Roeck and T. Gehrmann, hep-ph/9711512,
{\it in} Proc. Workshop on 
{\it Physics with Polarized Protons at HERA}, 
eds. A. De Roeck and T. Gehrmann, DESY-Proceedings-1998-01
%
\bibitem{elsa}
The PHOENICS Collaboration (A. Bock et al),
Phys. Rev. Lett. {\bf 81} (1998) 534.
%
\bibitem{mami}
The TAPS Collaboration (B. Krusche et al.),
Phys. Rev. Lett. {\bf 74} (1995) 3736.
%
\bibitem{cebaf}
CEBAF experiments
E-89-039 (S. Dytman et al.) and E-91-008 (B.G. Ritchie et al.).
%
\bibitem{hera}
S. Tapprogge, Talk at the Workshop {\it Pomeron and Odderon 
in Theory and Experiment} (Heidelberg, March 1998); \\
http://www.tphys.uni-heidelberg.de/ws/
%
\bibitem{celsius}
CELSIUS (H. Calen et al.), 
Phys. Rev. Lett. {\bf 80} (1998) 2069; \\
The WASA Collaboration (R. Bilger et al.), Nucl. Phys. {\bf A626} (1997) 93c.
%
\bibitem{cosy}
The COSY-11 Collaboration (P. Moskal et al.), 
Phys. Rev. Lett. {\bf 80} (1998) 3202.
%
\bibitem{wa102}
The WA102 Collaboration (D. Barberis et al.), 
Phys. Lett. {\bf B427} (1998) 398.
%
\bibitem{formf}
The CLEO Collaboration (J. Gronberg et al.),
Phys. Rev. {\bf D57} (1998) 33.
%
\bibitem{cleo}
The CLEO Collaboration (B.H. Behrens et al.),
Phys. Rev. Lett. {\bf 80} (1998) 3710; \\
The CLEO Collaboration (T.E. Browder et al.),
Phys. Rev. Lett. {\bf 81} (1998) 1786; \\
J.G. Smith, hep-ex/9803028.
%
\bibitem{fritzschb}
H. Fritzsch, Phys. Lett. {\bf B415} (1997) 83.
%
\bibitem{jafpl}
R.L. Jaffe, Phys. Lett. {\bf B365} (1996) 359.
%
\bibitem{man90}
A.V. Manohar, Phys. Rev. Lett. {\bf 65} (1990) 2511.
%
\bibitem{kogut}
J. Kogut and L. Susskind, Phys. Rev. {\bf D11} (1974) 3594.
%
\bibitem{cron}
C. Cronstr\"om and J. Mickelsson, J. Math. Phys. {\bf 24} (1983) 2528.
%
\bibitem{brandeis}
S.L. Adler, in Brandeis 
{\it Lectures on Elementary Particles and Quantum Field Theory}, 
eds. S. Deser, M. Grisaru and H. Pendleton (MIT Press, 1970).
%
\bibitem{crewther}
R.J. Crewther, Riv. Nuovo Cimento {\bf 2} (1979) 63, section 7.
%
\bibitem{bcs}
J. Bardeen, L. Cooper and J. Schrieffer, Phys. Rev. {\bf 108} (1957) 1175.
%
\bibitem{anderson}
P.W. Anderson and W.F. Brinkman, 
Theory of Anisotropic Superfluidity in He$^3$,
{\it in} 
The Helium Liquids: 
Proc. 15$^{th}$ Scottish Universities Summer School in Physics 1974, 
eds. J.G.M. Armitage and I.E. Farquhar (Academic Press, New York, 1975).
%
\bibitem{he4}
S. Grebenev, J.P. Toennies and A.F. Vilesov, Science {\bf 279} (1998) 2083.
%
\bibitem{svz}
  M.A. Shifman, A.I. Vainshtein and V.I. Zakharov,
  Nucl. Phys. {\bf B147} (1979) 385, 448; \\
  E.V. Shuryak, Phys. Rept. {\bf 115} (1984) 151.
%
\bibitem{bint}
S.D. Bass, B.L. Ioffe, N.N. Nikolaev and A.W. Thomas,
J. Moscow Phys. Soc. {\bf 1} (1991) 317.
%
\bibitem{mank}
L. Mankiewicz and A. Sch\"afer, Phys. Lett. {\bf B242} (1990) 455; \\
L. Mankiewicz, Phys. Rev. {\bf D43} (1991) 64.
%
\bibitem{bodwin}
G.T. Bodwin and J. Qiu, Phys. Rev. {\bf D41} (1990) 2755.
%
\bibitem{manoh}
A.V. Manohar, Phys. Rev. Lett. {\bf 66} (1991) 289.
%
\bibitem{ratc}
P. Ratcliffe, Nucl. Phys. {\bf B223} (1983) 45.
%
\bibitem{bnt}
S.D. Bass, N.N. Nikolaev and A.W. Thomas,
Adelaide University preprint ADP-133-T80 (1990) unpublished; \\
S.D. Bass, Ph.D. thesis (University of Adelaide, 1992).
%
\bibitem{bass92}
S.D. Bass, Z Physik {\bf C55} (1992) 653.
%
\bibitem{chengx}
H.-Y. Cheng, Phys. Lett. {\bf B427} (1998) 371.
%
\bibitem{strat}
M. Stratmann, hep-ph/9710379
{\it in} Proc. Workshop on {\it Deep Inelastic Scattering off Polarized 
Targets: Theory meets Experiment}, 
DESY-Zeuthen 1997, eds. J. Bl\"umlein et al. (DESY report 97-200, 1997).
%
\bibitem{grsv}
M. Gluck, E. Reya, M. Stratmann and W. Vogelsang,
Phys. Rev. {\bf D53} (1996) 4775.
%
\bibitem{eks}
J. Ellis, M. Karliner and C.T. Sachrajda, Phys. Lett. {\bf B231} (1989) 497.
%
\bibitem{lepage}
G.P. Lepage and S.J. Brodsky, Phys. Rev. {\bf D22} (1980) 2157.
%
\bibitem{bassbs}
S.D. Bass, S.J. Brodsky and I. Schmidt, Phys. Lett. {\bf B437} (1998) 417.
%
\bibitem{altp}
G. Altarelli and G. Parisi, Nucl. Phys. {\bf B126} (1977) 298.
%
\bibitem{neervan}
R. Mertig and W.L. van Neervan, Z Phys. {\bf C70} (1996) 637; \\
E.B. Zijlstra and W.L. van Neervan, Nucl. Phys. {\bf B417} (1994) 61;
(E) {\bf B426} (1994) 245.
%
\bibitem{vogelsang}
W. Vogelsang, Phys. Rev. {\bf D54} (1996) 2023.
%
\bibitem{leader}
E. Leader, A.V. Sidorov and D.B. Stamenov, Phys. Lett. {\bf B445} (1998) 232.
%
\bibitem{gehr}
T. Gehrmann and W.J. Stirling,
Phys. Rev. {\bf D53} (1996) 6100.
%
\bibitem{abfr}
G. Altarelli, R.D. Ball, S. Forte and G. Ridolfi, 
Nucl. Phys. {\bf B496} (1997) 337.
%
\bibitem{defl}
D. de Florian, O.A. Samapayo and R. Sassot, Phys. Rev {\bf D57} (1998) 5803.
%
\bibitem{ramsey}
L.E. Gordon, M. Goshtasbpour and G.P. Ramsey, 
Phys. Rev. {\bf D58} (1998) 094017.
%
\bibitem{e154qcd}
The E-154 Collaboration (K. Abe et al.), Phys. Lett. {\bf B405} (1997) 180.
%
\bibitem{thv}
G. 't Hooft and M. Veltman, Nucl. Phys. {\bf B44} (1972) 189.
%
\bibitem{forteab}
R.D. Ball, S. Forte and G. Ridolfi, 
Phys. Lett. {\bf B378} (1996) 255.
%
\bibitem{llew}
C.H. Llewellyn Smith, hep-ph/9812301.
%
\bibitem{compass} The COMPASS proposal, CERN/SPSLC 96-14.
%
\bibitem{hermesc}  The HERMES Charm Upgrade Program, HERMES 97-004.
%
\bibitem{bosted}
P. Bosted, private communication.
%
\bibitem{rhic}
N. Hayashi, Y. Goto and N. Saito, hep-ex/9807033.
%
\bibitem{heraproc}
Proc. Workshop on 
{\it Physics with Polarized Protons at HERA}, 
eds. A. De Roeck and T. Gehrmann, DESY-Proceedings-1998-01
%
\bibitem{semia}
L.L. Frankfurt et al., Phys Lett. {\bf B230} (1989) 141.
%
\bibitem{semib}
F.E. Close and R.G. Milner, Phys. Rev. {\bf D44} (1991) 3691.
%
\bibitem{semiexpt}
The SMC Collaboration (B. Adeva et al), Phys. Lett. {\bf B420} (1998) 180.
%
\bibitem{wislicki}
W. Wislicki, Mod. Phys. Lett. {\bf A13} (1988) 405.
%
\bibitem{kutiw}
J. Kuti and V. Weisskopf, Phys. Rev. {\bf D4} (1971) 3418.
%
\bibitem{closev}
F.E. Close, Nucl. Phys. {\bf B80} (1974) 269.
%
\bibitem{carlitz}
R. Carlitz and J. Kaur, Phys. Rev. Lett. {\bf 38} (1977) 673.
%
\bibitem{reinhardt}
H. Weigel, L. Gamberg and H. Reinhardt, Phys. Rev. {\bf D55} (1997) 6910; \\
O. Schroder, H. Reinhardt and H. Weigel, Phys. Lett. {\bf B439} (1998) 398.
%
\bibitem{brodbs}
S.J. Brodsky, M. Burkardt and I. Schmidt, Nucl. Phys. {\bf B441} (1995)
197.
%
\bibitem{e155}
C. Young
{\it in} Proc. Workshop on {\it Deep Inelastic Scattering off Polarized 
Targets: Theory meets Experiment}, 
DESY-Zeuthen 1997, eds. J. Bl\"umlein et al. (DESY report 97-200, 1997).
%
\bibitem{soff}
J. Soffer and O.V. Teryaev, Phys.Rev. {\bf D56} (1997) 1549.
%
\bibitem{mart}
S.D. Bass and M.M. Brisudov{\'a}, 
hep-ph/9711423, Euro. Phys. J {\bf A} (in press).
%
\bibitem{heim}
R.L. Heimann, Nucl. Phys. {\bf B64} (1973) 429.
%
\bibitem{ek}
J. Ellis and M. Karliner, Phys. Lett. {\bf B213} (1988) 73.
%
\bibitem{pvl1}
P.V. Landshoff, 
Proc. Zuoz Summer School, PSI Proceedings 94-01 (1994) 135,
hep-ph/9410250.
%
\bibitem{nmcf2}
The New Muon Collaboration (M. Arneodo et al.), 
Nucl. Phys. {\bf B483} (1997) 3.
%
\bibitem{badk}
B. Bade{\l}ek and J. Kwieci{\'n}ski, Phys. Lett. {\bf B418} (1998) 229.
%
\bibitem{zeut}
Various contributions {\it in} Proc. Workshop on {\it Deep Inelastic
Scattering off Polarized Targets: Theory meets Experiment}, 
DESY-Zeuthen 1997, eds. J. Bl\"umlein et al. (DESY report 97-200, 1997).
%
\bibitem{kuti}
L. Galfi, J. Kuti and A. Patkos, Phys. Lett. {\bf B31} (1970) 465; \\
J. Kuti, Erice lectures (1995), 
{\it in} Proc. Erice School {\it The spin structure of the nucleon}, 
eds. B. Frois and V. Hughes (World Scientific, 1997).
%
\bibitem{clos}
F.E. Close and R.G. Roberts, Phys. Rev. Lett. {\bf 60} (1988) 1471; 
Phys. Lett. {\bf B336} (1994) 257.
%
\bibitem{sbpl}
S.D. Bass and P.V. Landshoff, Phys. Lett. {\bf B336} (1994) 537.
%
\bibitem{kirs}
R. Kirschner and L.N. Lipatov, Nucl. Phys. {\bf B213} (1983) 122.
%
\bibitem{bart}
J. Bartels, B.I. Ermolaev and M.G. Ryskin, 
Z Phys {\bf C70} (1996) 273; {\bf C72} (1996) 627.
%
\bibitem{blum}
J. Bl{\"u}mlein and A. Vogt, Phys. Lett. {\bf B386} (1996) 350.
%
\bibitem{gott}
K. Gottfried, Phys. Rev. Lett. {\bf 18} (1967) 1174.
%
\bibitem{nmc}
The New Muon Collaboration 
(M. Arneodo et al.), Phys. Rev. {\bf D50} (1994) R1.
%
\bibitem{nusea}
The E866/NuSea Collaboration (E.A. Hawker et al.), 
Phys. Rev. Lett. {\bf 80} (1998) 3715.
%
\bibitem{pion}
A.W. Thomas, Phys. Lett. {\bf B126} (1983) 97.
%
\bibitem{feynman}
R.D. Field and R.P. Feynman, Phys. Rev. {\bf D15} (1977) 2590.
%
\bibitem{shimoda}
A. W. Thomas and W. Melnitchouk, in {\it New  Frontiers in Nuclear
Physics},
eds. S. Homma, Y. Akaishi and M. Wada (World Scientific, Singapore, 1993),
pp. 41-106.
%
\end{thebibliography}
\end{document}